\documentclass[aps,prd,twocolumn,nofootinbib,longbibliography,superscriptaddress,10pt]{revtex4-2}
\usepackage{etoolbox}
\usepackage{dcolumn}
\usepackage{amsmath,amssymb,amsfonts,mathtools}
\usepackage{mathrsfs}
\usepackage{graphicx}
\usepackage[colorlinks=true,urlcolor=blue,citecolor=red,linkcolor=blue]{hyperref}
\usepackage{natbib}
\usepackage{wrapfig}
\usepackage{flushend}

\begin{document}
\title{\bf{\Large  Thermodynamics of a Schwarzschild black hole surrounded by quintessence in the generalized uncertainty principle framework}}
\author{Soham Sen}
\email{sensohomhary@gmail.com}
\affiliation{Department of Astrophysics and High Energy Physics, S. N. Bose National Centre for Basic Sciences, JD Block, Sector-III, Salt Lake City, Kolkata-700 106, India}
\author{Abhijit Dutta}
\email{dutta.abhijit87@gmail.com,abhijit.dutta@krc.edu.in}
\affiliation{Department of Physics, Kandi Raj College, Kandi, Murshidabad-742 137, India}
\author{Sunandan Gangopadhyay}
\email{sunandan.gangopadhyay@gmail.com}
\affiliation{Department of Astrophysics and High Energy Physics, S. N. Bose National Centre for Basic Sciences, JD Block, Sector-III, Salt Lake City, Kolkata-700 106, India}
%%%%%%%%%%%%%%%%%%%%

\begin{abstract}
\noindent We investigate the thermodynamics of a Schwarzschild black hole, surrounded by the quintessence energy-matter in the linear and quadratic generalized uncertainty principle framework. Considering the variance in the position to be of the order of the event horizon radius and equating the variance in the momentum to the Hawking temperature of the black hole, we substitute these variances in the deformed algebra. From there we obtained the generalized uncertainty principle-modified black hole temperature and eventually the specific heat of the black hole. Then we calculate the critical as well as the remnant mass and obtain the entropy relation. We observe that the entropy relation includes the usual leading order ``\textit{area divided by four}" term, sub-leading logarithmic term, and higher order inverse of the area corrections. Finally, calculating the energy output as a function of time, we obtain the evaporation time of the black hole. The results show the dependence of the quintessence parameter on the thermodynamic quantities in the framework of linear and quadratic generalized uncertainty principle.
\end{abstract}

\maketitle
%%%%%%%%%%%%%%%%%%%%%%%%%%%%%%%%%%%%%%%%%%%%%%%

\section{Introduction}
\noindent The existence of black holes in the general theory of relativity is considered as one of the biggest theoretical predictions in the history of physics. With the imaging of the shadow of the black hole Sagittarius A*, black holes are proved to be real astronomical objects in existence with their bizarre nature. One of the most important aspects of black holes is their thermodynamic behaviour which is yet to be verified. A series of papers by Bekenstein \cite{Bekenstein,Bekenstein2} and Hawking \cite{Hawking,Hawking2,Hawking3} had shown that a black hole can be considered as a thermodynamical object where the entropy of the black hole depends on the area of the black hole. The prediction specifically claimed that black holes are not as black as they are thought to be and the radiation was termed as the Hawking radiation \cite{Hawking,Hawking2,Hawking3}. This relation between black holes and thermodynamics has been investigated in several analyses over a long period \cite{CalCogCle,Page,JacWall,Zhang,
AppGreNak,Astorino,Bayraktar,Dehghani,HuOngPage,
YaoHouOng,KrtousZelnikov,DuttaPadhyay,
DuttaPadhyay2,DuttaPadhyay3,DuttaPadhyay4,
DuttaPadhyay5,DuttaPadhyay6}. One of the propositions regarding our universe is that it is continuously expanding \cite{Riess0,Riess1} and it was explained by considering the existence of dark energy. Although dark energy is not detected still today, it would be much more prudent to investigate black hole thermodynamics in the presence of dark energy. It is considered that if a black hole is surrounded by dark energy, it will have important effects on the thermodynamics of a black hole. In \cite{Kiselev}, the thermodynamics of a Schwarzschild black hole surrounded by quintessence matter. Quintessence matter is basically one of the dark energy candidates. In \cite{Chen}, the Hawking radiation for a $d$-dimensional spherically symmetric black hole surrounded by quintessence matter has been investigated. Later on people have investigated the thermodynamics of Reissner-Nordstr\"{o}m \cite{RN1,RN2}, Narai type \cite{Narai1,Narai2}, Bardeen \cite{Bardeen1,Bardeen2} and Schwarzschild black holes with quantum corrections \cite{SchQM}.

\noindent Although the general theory of relativity is the most accurate classical field theory, a quantum mechanical version of it is still not unveiled. People have tried to put forward several quantum gravity theories like string theory \cite{String0,String1,String2}, loop quantum gravity \cite{lqg1,lqg2,lqgRovelli}, and non-commutative geometry \cite{ncg1,ncgconnes} but none of them have been quite successful in giving an ultimate description of physics at or near the Planck length. One thing that is quite clear from most of the analyses is that there exists a fundamental length scale in nature. The easiest way to incorporate such fundamental length scales in a quantum mechanical description is to modify Heisenberg's uncertainty principle. The Heisenberg's uncertainty principle in one-dimensional spacetime takes the form \cite{Kempf}\footnote{It is important to note that we are in the geometrized units, i.e. $c=G=1$.}
\begin{equation}\label{1.1}
\Delta x\Delta p\geq \frac{\hbar}{2}\left(1+\frac{\beta^2 l_p^2}{\hbar^2}(\Delta p)^2\right)~,
\end{equation}
where $\beta$ is a dimensionless constant and $l_p=\sqrt{\frac{\hbar G}{c^3}}$ is the Planck length\footnote{We shall also like to mention some articles involving the GUP effect in various physical phenomena \cite{GUP1,GUP2,GUP3,GUP3B,GUP4,GUP5}}. In \cite{OTMGraviton}, a detector-graviton interaction model in the presence of Earth's gravity has been used which has led to an uncertainty principle induced by the noise of gravitons. This uncertainty relation obtained in \cite{OTMGraviton} reduces to the uncertainty principle in eq.(\ref{1.1}) in the Planck mass limit. On the other hand, doubly special relativity (DSR) theories \cite{DSR1,DSR2,DSR3} suggest the existence of a maximum observable momentum along with the fundamental minimal length scale which results in a linear order modification in the momentum uncertainty along with that of the quadratic order modification term in eq.(\ref{1.1}). The DSR modified uncertainty principle reads \cite{LQGUP1,LQGUP2,LQGUP3} 
\begin{equation}\label{1.2}
\Delta x\Delta p\geq \frac{\hbar}{2}\left(1-\frac{\alpha l_p}{\hbar}\Delta p+\frac{\beta^2 l_p^2}{\hbar^2}(\Delta p)^2\right)~,
\end{equation}
where $\alpha$ and $\beta$ are dimensionless constants.
The above uncertainty relation is also known as the linear-quadratic uncertainty principle or LQGUP. Thermodynamics of a black hole in the presence of a generalized uncertainty principle framework has been studied extensively in \cite{BHThermo1,BHThermo2,BHThermo3,
BHThermo4,BHThermo5,BHThermo6,
BHThermo7,BHThermo8,BHThermo9,
BHThermo10,BHThermo11,BHThermo12,
BHThermo13,BHThermo14,BHThermo15,
BHThermo16,BHThermo17,BHThermo18,
BHThermo19,BHThermo20,BHThermo21}. One of the most important effects of generalized uncertainty principle framework in black hole thermodynamics is that the GUP correction prevents the black hole from completly evaporating away. To construct the uncertainty relation in terms of black hole parameters, in general, one identifies the event horizon of the black hole with the uncertainty in measurement of the position variable whereas the Hawking temperature is related to the uncertainty in the momentum variable. Using this identification, one can construct a mass-temperature relation by substituting $\Delta x$ and $\Delta p$ in eq.(\ref{1.2}). It is important to observe that only the equality condition in eq.(\ref{1.2}) is used to construct the mass-temperature relation. Using this relation, one calculate all of the important thermodynamical quantites of the black hole which are usually expressed in terms of its event horizon radius. Recently in \cite{QuintSchwarzQGUP}, the authors have calculated the thermodynamics of a Schwarzschild black hole surrounded by quintessence matter in the generalized uncertainty principle framework where the uncertainty in the position and momentum variables follow eq.(\ref{1.1}). In our current analysis, we have extended the analysis presented in \cite{QuintSchwarzQGUP} by considering the LQGUP framework. Following the standard procedure, we obtain the mass-temperature relation in the LQGUP framework. Then we have calculated the critical mass, specific heat, remnant mass, entropy, and the energy density of a Schwarzschild black hole surrounded by quintessence matter.

\noindent Our paper is organized as follows. In section (\ref{S2}), we discuss the Schwazschild black hole surrounded by quintessence matter and obtain the event horizon radii for fixed values of the quintessence parameter $\omega_q$. Then in section (\ref{S3}), we discuss regarding the thermodynamics of the black hole in the LQGUP framework. In section (\ref{S4}), by considering energy output as a function of time, we obtain the evaporation equation and eventually the evaporation time. In section (\ref{S4B}), we discuss the key differences of our analysis with the existing works on black hole analysis and finally conclude in section (\ref{S5}).
%%%%%%%%%%%%%%%%%%%%%%%%%%%%%%%%%%%%%%%%%%%%%%%%%%%%%%%

\section{Schwarzschild black hole fenced in quintessence}\label{S2}
\noindent We start our analysis by considering a static spherically symmetric black hole of mass $M$ surrounded by the quintessence energy-matter. The exact solution of the Einstein field equation for this black hole has been derived in \cite{Kiselev} and is generally known as the ``\textit{Kiselev}" solution. The general form of the metric reads \cite{Kiselev}
\begin{equation}
	ds^2=-f(r)dt^2+\frac{1}{f(r)}dr^2+r^2 d{\Omega_2}^2~,
	\label{2.1}
\end{equation} 
where the lapse function $f(r)$ takes the form
\begin{equation}
	f(r)=1-\frac{2M}{r}-\frac{\eta_{\omega_q}}{r^{3\omega_q+1}}~.
	\label{2.2}
\end{equation}
In the above equation, $\omega_q$ is known as the quintessential state parameter and $\eta_{\omega_q}$ is the positive normalization factor depending on the density of the quintessence matter. It is quite well-known that in the range $-1 <\omega_q < -\frac{1}{3}$, the quintessence effortlessly intercepts the accelerated expansion of the universe. It is evident from the form of $f(r)$ in eq.(\ref{2.2}) that it reduces to the usual Schwarzschild metric in the $\eta_{\omega_q}\rightarrow 0$ limit. In the vicinity of the quintessence matter, the nonvanishing components of the mixed energy-momentum tensor $T^{~\mu}_{\nu}$ read
\begin{align}
	{T_t}^t&={T_r}^r=-\rho_q\label{2.3}\\
	{T_\theta}^\theta&={T_\phi}^\phi=\frac{\rho_q}{2}(3\omega_q+1)\label{2.4}
\end{align}
with the matter density $\rho_q$ given by
\begin{equation}\label{2.5}
\rho_q = -\frac{3}{2} \frac{\eta_{\omega_q}\omega_{q}}{r^{3(\omega_q+1)}}~.
\end{equation}
The pressure $P_q$ in terms of the matter density $\rho_q$ reads
\begin{equation}\label{2.6}
	 P_q= \omega_q \rho_q~.
\end{equation}
To avoid cluttering the fundamental constants in the results, we work in the geometrized units which implies $c=G=1$. We also set the Boltzmann constant $k_B$ to unity. In order to obtain the event horizon radii, we make use of the vanishing condition for the lapse function $f(r)$ as
\begin{equation}\label{2.7}
	f(r)| _{r=r_H} =0
\end{equation}
from which, by using eq.(\ref{2.2}), we get
\begin{equation}\label{2.8}
	r_H-\frac{\eta_{\omega_q}}{{r_H}^{3\omega_q}}=2M~.
	\end{equation}
Following \cite{QuintSchwarzQGUP}, we shall make use of three values of the quintessential state parameter $\omega_q$ given by $\omega_q=\{-\frac{2}{3},-\frac{1}{3},-1\}$. 
\begin{enumerate}

 \item For $\omega_q=-\frac{2}{3}$, two event horizon radii emerge, namely, inner horizon radius ($r_{H_{\text{in}}}$) or the Cauchy horizon of the black hole and the outer horizon radius $r_{H_{\text{out}}}$ or the event horizon radius of the black hole which are given by
\begin{align}
	r_{H_{\text{in}}}&=\frac{1-\sqrt{1-8\eta_{-2/3} M}}{2 \eta_{-2/3}}\nonumber\\
	r_{H_{\text{out}}}&=\frac{1+\sqrt{1-8\eta_{-2/3} M}}{2 \eta_{-2/3}}~.\label{2.9}
\end{align}
\noindent The notable point is that the outer horizon is commonly known as the quintessence horizon identical to the cosmological horizon in de Sitter spacetime.

\item For $\omega_q=-\frac{1}{3}$, we get only one event horizon radius which reads
\begin{equation}\label{2.10}
	r_H = \frac{2M}{1-\eta_{-1/3}}~.
\end{equation}
\item For $\omega_q=-1$, we have a perturbative solution for the horizon radius as
\begin{equation}\label{2.11}
r_H\simeq 2M \left(1+4\eta_{-1} M^2\right)~.
\end{equation}
\end{enumerate}

\section{Thermodynamics of Black Hole}\label{S3}
\noindent To investigate the thermodynamics of a black hole, a particle-anti-particle pair production in the vicinity of the event horizon of the black hole is considered.  The particle with negative energy falls inside the outer horizon (or the event horizon) of the black hole and that with the positive energy escapes outside the event horizon and gets observed by an asymptotic observer. The particle is considered to be massless. It can be easily inferred that the momentum $p$ of the particle emitted outside of the event horizon of the black hole characterizes the temperature $T$ of the black hole. The temperature can be considered to be proportional to the uncertainty in the momentum $\Delta p$ of the emitted particles. Hence, one can express $T$ as \cite{Adler}
\begin{equation}\label{2.12}
	T=\frac{c \Delta p}{k_B}
	\end{equation}
where $c$ is the speed of light and $k_B$ is the Boltzmann constant. Following our choice of constants, one can reinterpret the temperature in eq.(\ref{2.12}) as $T=\Delta p$. The Hawking temperature of the black hole will be equal to the temperature of the particle when thermodynamic equilibrium is reached. The uncertainty in the position of a particle near the event horizon of the Schwarzschild black hole will be of the order of the Schwarzschild radius of the black hole \cite{Adler,Medved}
\begin{equation}\label{2.13}
	\Delta x=\epsilon r_{H}
\end{equation}
where $\epsilon$ is an undetermined constant and $r_H$ is the event horizon radius of the black hole. In order to fix the undetermined constant in eq.(\ref{2.13}), we consider the equality condition in eq.(\ref{1.2}) in the Heisenberg uncertainty  limit ($\alpha,\beta\rightarrow 0$) as 
\begin{equation}\label{2.14}
\begin{split}
\Delta x \Delta p&=\frac{\hbar}{2}\\
\implies \epsilon r_{H} T_{\text{HUP}}&=\frac{\hbar}{2}~.
\end{split}
\end{equation}
We next need to consider the case when the black hole is not surrounded by the quintessence energy matter and in this $\eta_{\omega_q}\rightarrow 0$ limit, $r_H=2M$ comes out from eq.(\ref{2.8}). In this limit one can recast eq.(\ref{2.14}) as
\begin{equation}\label{2.15}
T_{\text{HUP}}=\frac{\hbar}{4\epsilon M}~.
\end{equation}
Now, the Hawking temperature of the black hole can be expressed in terms of its surface gravity $\kappa$ as
\begin{equation}\label{2.16}
T_{H}=\frac{\hbar \kappa}{2\pi}
\end{equation}
where $\kappa=\frac{\partial_r f(r)\rvert_{r=r_H}}{2}$. In the $\eta_{\omega_q}\rightarrow 0$ limit, eq.(\ref{2.16}) reduces to
\begin{equation}\label{2.17}
T_H=\frac{\hbar}{8\pi M}~.
\end{equation}
Equating eq.(\ref{2.17}) with eq.(\ref{2.15}), we obtain the value of the constant $\epsilon$ to be $2\pi$. Our next aim is to express the Hawking temperature of the black hole in terms of the horizon radius (mass) of the black hole.

\noindent As we are in the geometrized units, the relations $\frac{c\hbar}{l_p}=m_{p}c^{2}$ and $m_{p}=\frac{c^2 l_p}{G}$  ($m_{p}$ being the Planck mass) can be modified as 
\begin{equation}\label{2.18}
	\frac{\hbar}{l_p}=m_{p}~, m_{p}=l_p~.
\end{equation}
 The equality condition from the LQGUP relation in eq.(\ref{1.2}) can be recast as
 \begin{equation}\label{2.19}
 \begin{split}
 \Delta x\Delta p&=\frac{\hbar}{2}\left(1-\frac{\alpha}{m_p}\Delta p+\frac{\beta^2}{m_p^2}(\Delta p)^2\right)\\
 \implies 2\pi r_H T&=\frac{\hbar}{2}\left(1-\frac{\alpha}{m_p}\Delta p+\frac{\beta^2}{m_p^2}(\Delta p)^2\right)\\
 \implies r_H&=\frac{\hbar}{4\pi T}\left(1-\frac{\alpha}{m_p}T+\frac{\beta^2}{m_p^2}T^2\right)~.
 \end{split} 
 \end{equation}
From the above relation, one can write down the GUP corrected temperature in terms of the event horizon radius as
\begin{equation}\label{2.20}
\begin{split}
T_\pm=&\frac{1}{2\beta^2}\left(\alpha m_p+4\pi r_H\right)\\&\pm\frac{1}{2\beta^2}\sqrt{\left(\alpha m_p+4\pi r_H\right)^2-4m_P^2\beta^2}~,
\end{split}
\end{equation}
where the minus sign is taken to be the Hawking temperature $T=T_-$. In the next few subsections, we shall calculate the critical mass, specific heat, remnant mass, entropy, energy density, and the pressure of the black hole considering three fixed values of the quintessential state parameter.
% \noindent Now we will simplify temperature ($T$) in eq. (\ref{Temp. in terms of horizon radius}) as 
% \begin{eqnarray}
 %	T&&= \frac{m_p}{2 \hbar \gamma^2} (4 \pi m_p r_H + \hbar \beta) \left[1-\sqrt{1 - \frac{4 \hbar^2 \gamma^2}{(4 \pi m_p r_H + \hbar \beta)^2}}\right] \nonumber\\
% 	&&= \frac{\hbar m_p}{R}\left[1+\frac{\hbar^2 \gamma^2}{R^2} + \frac{2 \hbar^4 \gamma^4}{R^4}\right]~, 
% 	\label{simplified temp. in terms of horizon radius}
 %\end{eqnarray}
 %\noindent  where, $R =4 \pi m_p r_H + \hbar \beta $ has been used.
 \subsection{Critical Mass}
 \noindent  In this subsection, we shall calculate the critical mass of the black hole, below which the thermodynamic quantities become ill-defined. In order for the temperature to be real-valued in eq.(\ref{2.20}), the term inside the square root must be equal to or greater than zero. Hence, from eq.(\ref{2.20}), we obtain the following inequality for a real-valued temperature as
\begin{equation}\label{2.21}
\begin{split}
&(\alpha m_p+4\pi r_H)^2-4 m_p^2 \beta^2\geq 0\\
\implies &r_{H_{\text{Cr.}}}=\frac{(2\beta-\alpha)m_p}{4\pi}~.
\end{split}
\end{equation}
The above equation gives the value of the critical radius ($r_{H_{\text{Cr.}}}$) below which the temperature becomes complex-valued. From the last line of the above equation, we shall calculate the critical mass for three cases.
\begin{enumerate}

\item For  $\omega_q=-2/3$, we know from eq.(\ref{2.9}) that there are two values of the horizon radius. As a result, one will get two values of the critical mass corresponding to the two horizon radii. 
\begin{enumerate}
\item Firstly, for ${r_{H_{\text{in}}}}$, we get
 \begin{equation}\label{2.22}
 \begin{split}
 	&\frac{1-\sqrt{1-8 \eta_{-2/3} M_{\text{Cr.}_{\text{in}}}}}{2 \eta_{-2/3}} = \frac{m_p}{4 \pi}(2\beta - \alpha)\\
 	\implies &M_{\text{Cr.}_{\text{in}}}= \frac{m_p}{8 \pi}(2\beta - \alpha) - \frac{\eta_{-2/3}m_p^2}{32 \pi^2}(2\beta - \alpha)^2~.
 	\end{split}
 \end{equation}
 
 \item Secondly, for $r_{H_{out}}$, we get
 \begin{equation}\label{2.23}
 \begin{split}
 	&\frac{1+\sqrt{1-8 \eta_{-2/3} M_{\text{Cr.}_{\text{out}}}}}{2 \eta_{-2/3}} = \frac{m_p}{4 \pi}(2\beta - \alpha)\\
 	\implies &M_{\text{Cr.}_{\text{out}}}= \frac{m_p}{8 \pi}(2\beta - \alpha) - \frac{\eta_{-2/3}m_p^2}{32 \pi^2}(2\beta - \alpha)^2~.
 	\end{split}
 \end{equation}
 \noindent We see from eq.(s)(\ref{2.22},\ref{2.23}) that the critical masses of the black hole for both the cases (${r_{H_{\text{in}}}}$ and $r_{H_{\text{out}}}$ ) are same. This is an expected feature as the two radii corresponds to two different horizons of a single black hole with a fixed mass. As a result, the critical mass should be the same for a single black hole.
 \end{enumerate}
\item To find out the critical mass for the $\omega_q=-1/3$ case, we use the form of the horizon radius from eq.(\ref{2.10}) in the left-hand side of eq. (\ref{2.21}) and obtain
 \begin{equation}
 \begin{split}
 	\frac{2M_{\text{Cr.}}}{1-\eta_{-1/3}} &= \frac{m_p}{4 \pi }(2\beta - \alpha)\\
 \implies {M_{\text{Cr.}_{(\omega_q=-1/3)}}}& = \frac{m_p}{8 \pi }(1-\eta_{-1/3})(2\beta - \alpha)~.
 	\label{2.24} 
 	\end{split}
\end{equation}
 
\item We can find out the critical mass of the black hole for $\omega_q=-1$ from the vanishing condition of the lapse function much more easily. Instead of making use of the horizon radius in terms of the mass of the black hole, we use the $f(r_H)=0$ condition as follows
\begin{equation}\label{2.25}
\begin{split}
f(r_H)&=0=1-\frac{2M}{r_H}-\eta_{-1}r_H^2\\
\implies M&=\frac{r_H}{2}-\frac{\eta_{-1}r_H^3}{2}~.
\end{split}
\end{equation}
Now substituting the critical radius from eq.(\ref{2.21}), we arrive at the critical mass of the black hole to be
\begin{equation}\label{2.26}
M_{\text{Cr.}_{\omega_q=-1}}= \frac{m_p}{8 \pi }(2\beta - \alpha)-\frac{\eta_{-1}m_p^3}{128\pi^3}(2\beta-\alpha)^3~.
\end{equation}
In the absence of the quintessence energy matter, all the critical mass values in eq.(s)(\ref{2.22},\ref{2.23},\ref{2.24},\ref{2.26}) reduces to the same value $\frac{m_p}{8\pi}(2\beta-\alpha)$ which gives the critical mass value for the Schwarzachild black hole in the LQGUP framework when the quintessence energy matter is absent. 
\end{enumerate}

\subsection{Specific Heat}
\noindent In this subsection, we shall calculate the specific heat of the black hole in terms of its event horizon radius. The form of the specific  heat of the black hole from the first law of black hole thermodynamics reads
\begin{equation}\label{2.27}
\begin{split}
C&=\frac{dM}{dT}\\
\implies C&=\frac{dM}{dr_H}\frac{dr_H}{dT}~.
\end{split}
\end{equation}
It is evident from the above equation that we need to calculate two separate quantities to calculate the specific heat of the black hole. The second quantity which is the total derivative of the horizon radius in terms of the temperature of the black hole can be calculated directly from eq.(\ref{2.19}) and reads
\begin{equation}\label{2.28}
\frac{dr_H}{dT}=\frac{\hbar^2}{4\pi}\left(-\frac{1}{T^2}+\frac{\beta^2}{m_p^2}\right)~.
\end{equation}
The above quantity is the same for all of the three cases discussed earlier as it is not dependent on the $\eta_{\omega_q}$ parameter.
The second quantity $\frac{dM}{dr_{H}}$ is calculated by using the mass-event horizon relation obtained from the vanishing condition of the lapse function $f(r_H)=0$. This quantity will be different for different values of the quintessential state parameter.
\begin{enumerate}
\item For $\omega_q=-2/3$, we obtain the  mass-horizon radius relation as
\begin{equation}\label{2.29}
\begin{split}
f(r_H)&=0\\
\implies M&=\frac{r_H}{2}-\frac{\eta_{-2/3r_H^2}}{2}\\
\implies \frac{dM}{dr_H}&=\frac{1-2\eta_{-2/3}r_H}{2}~.
\end{split}
\end{equation}
Substituting eq.(\ref{2.29}) and eq.(\ref{2.28}) back in eq.(\ref{2.27}), one can obtain the form of the specific heat of the black hole up to $\mathcal{O}(\alpha^4,\beta^4)$\footnote{It is important to remember that $\alpha$ and $\beta$ are of the same order. } as
\begin{equation}\label{2.30}
\begin{split}
C_{r_{H_{\text{in}}}}=C_{r_{H_{\text{out}}}}&=\frac{\hbar}{8\pi}(1-2\eta_{-2/3}r_H)\left(-\frac{1}{T^2}+\frac{\beta^2}{m_p^2}\right)\\
\implies C_{(\omega_q=-2/3)}&\simeq-2\pi r_H^2(1-2\eta_{-2/3}r_H)\Bigr(1+\frac{\alpha m_p}{2\pi r_H}\\&+\frac{(\alpha^2-3\beta^2)m_p^2}{16\pi^2 r_H^2}-\frac{\beta^4 m_p^4}{256 \pi^4 r_H^4}\Bigr)~.
\end{split}
\end{equation}
We shall now plot the specific heat of the black hole against its event horizon radius. At first, we shall consider that the uncertainty product in eq.(\ref{1.2}) to be exact, which implies that we should use $T_{-}$ from eq.(\ref{2.20}) and substitute it back directly in the first line of eq.(\ref{2.30}). For plotting, we express the specific heat of the black hole from the first line of eq.(\ref{2.30}) in terms of its mass and compare it with the approximate result in the final line of eq.(\ref{2.30}) in terms of the mass of the black hole. It is crucial to remember that $\eta_{-2/3}r_H$ is a dimensionless and very small quantity, so we need to probe the regime with $r_H$ values for a fixed value of $\eta_{-2/3}$ such that the $\eta_{-2/3}r_H\ll1$. For plotting purposes, we have used $\hbar=1$ along with $\alpha=0.05,~\beta=0.05$ and $\eta_{-2/3}=0.01$ in Fig.(\ref{Fig1}).
\begin{figure}[ht!]
\begin{center}
\includegraphics[scale=0.28]{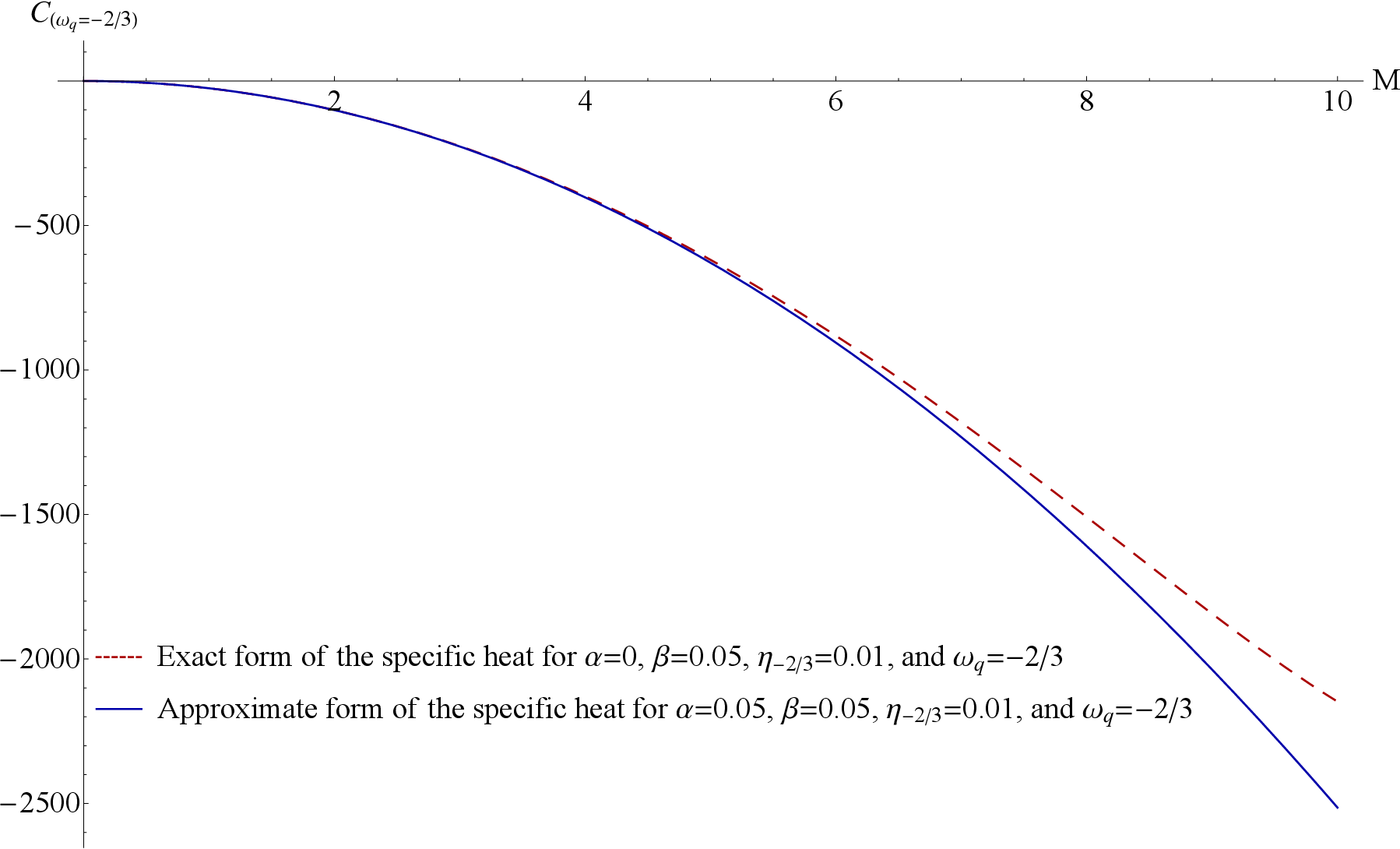}
\caption{Plot of the specific heat of the black hole against its mass when exact and approximated forms are used. Here the parameter values are set to $\alpha=0.05$, $\beta=0.05$, and $\eta=0.01$.\label{Fig1}}
\end{center}
\end{figure}
It is important to observe from Fig.(\ref{Fig1}) that as the mass approaches $10m_p$ value ($m_p=1$ when $\hbar=G=c=1$) then the $\eta_{-2/3}r_H\sim0.1$. As a result the approximate solution starts deviating from the ``exact" one. This implies that for black holes with smaller masses, the approximation used is quite appropriate. We shall now investigate the contribution of the linear GUP parameter towards the specific heat of the black hole. 
\begin{figure}[ht!]
\begin{center}
\includegraphics[scale=0.28]{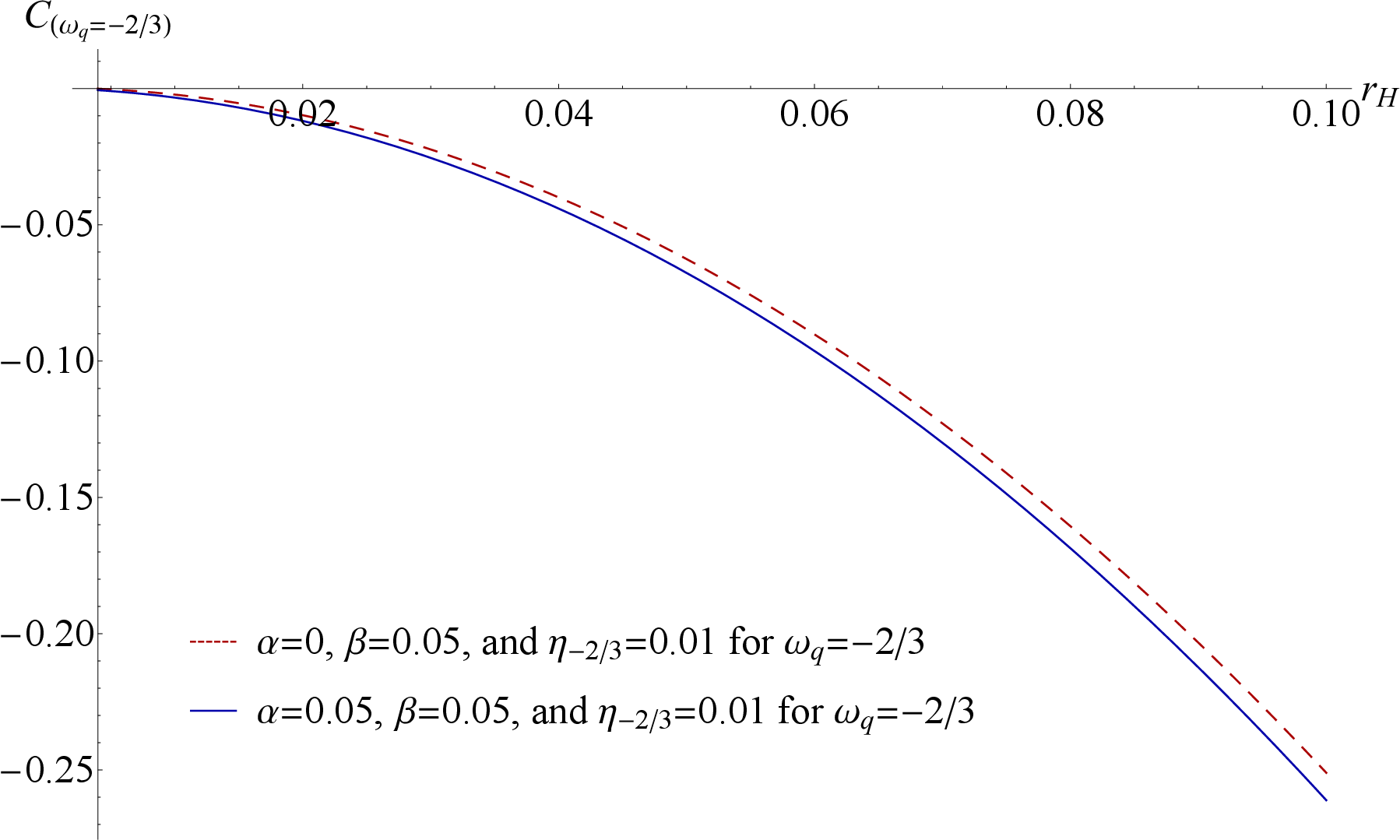}
\caption{Plot of the specific heat of the black hole against its event horizon radius when the linear GUP parameter ($\alpha$) is zero and not zero.\label{Fig2}}
\end{center}
\end{figure}
We have set $\beta=0.05$ and $\eta=0.01$ in Fig.(\ref{Fig2}). Comparing Fig.(\ref{Fig1}) and Fig.(\ref{Fig2}), one can easily find out that the specific heat drops faster with $r_H$ when the value of $\alpha$ is non-zero than the $\alpha=0$ case. We can further improvise the nature of the specific heat in Fig.(\ref{Fig3}).
\begin{figure}[ht!]
\begin{center}
\includegraphics[scale=0.28]{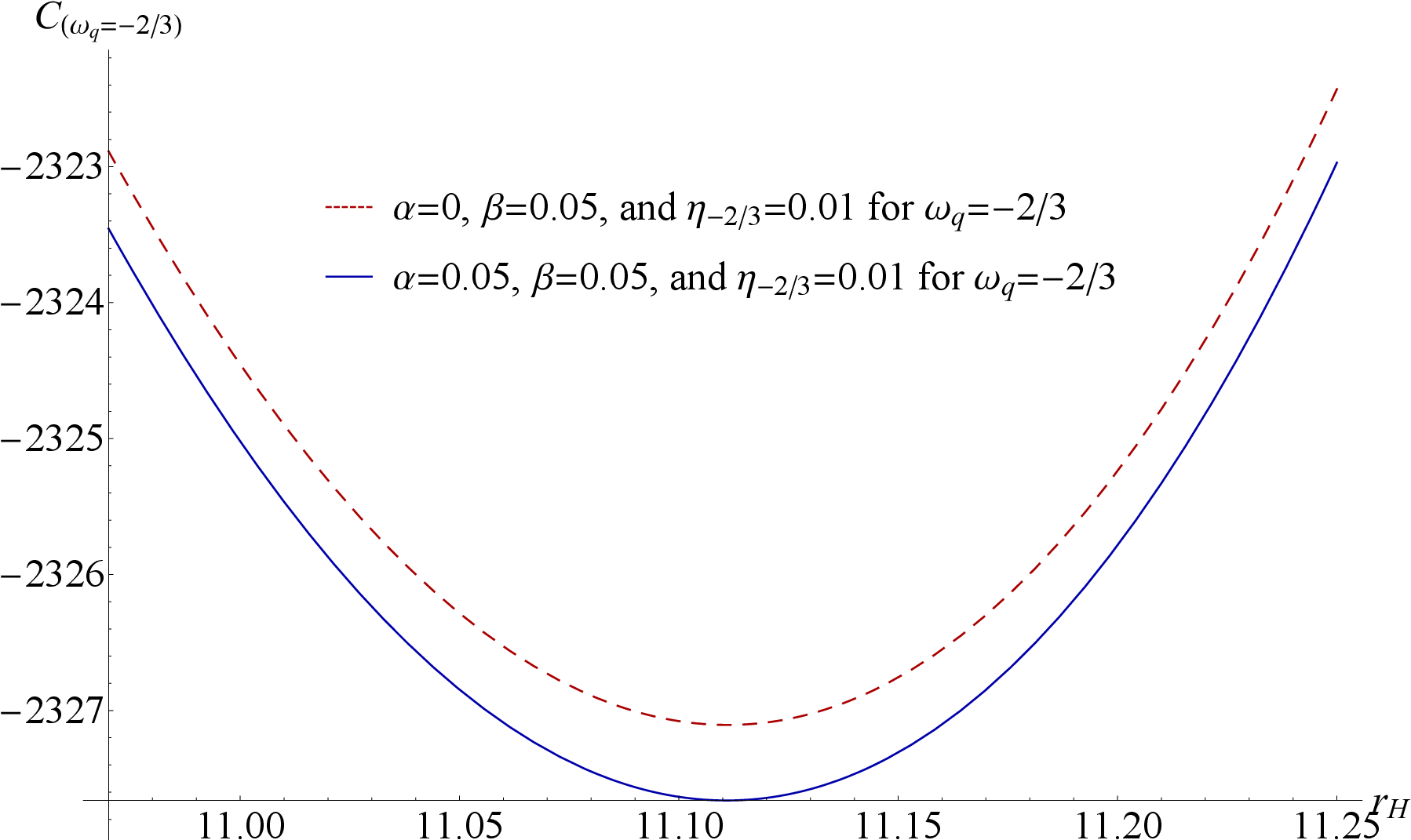}
\caption{Specific heat of the black hole against its event horizon radius is plotted when $\alpha=0$ and $\alpha=0.05$.\label{Fig3}}
\end{center}
\end{figure}
We can see that the specific heat decreases faster and then increases slowly with increasing value of the event horizon radius with $\alpha=0.05$ when plotted against the $\alpha=0$ case. Now, we plot the specific heat of the black hole for different values of the quintessential positive normalization factor $\eta_{-2/3}$ in Fig.(\ref{Fig4}).
\begin{figure}[ht!]
\begin{center}
\includegraphics[scale=0.28]{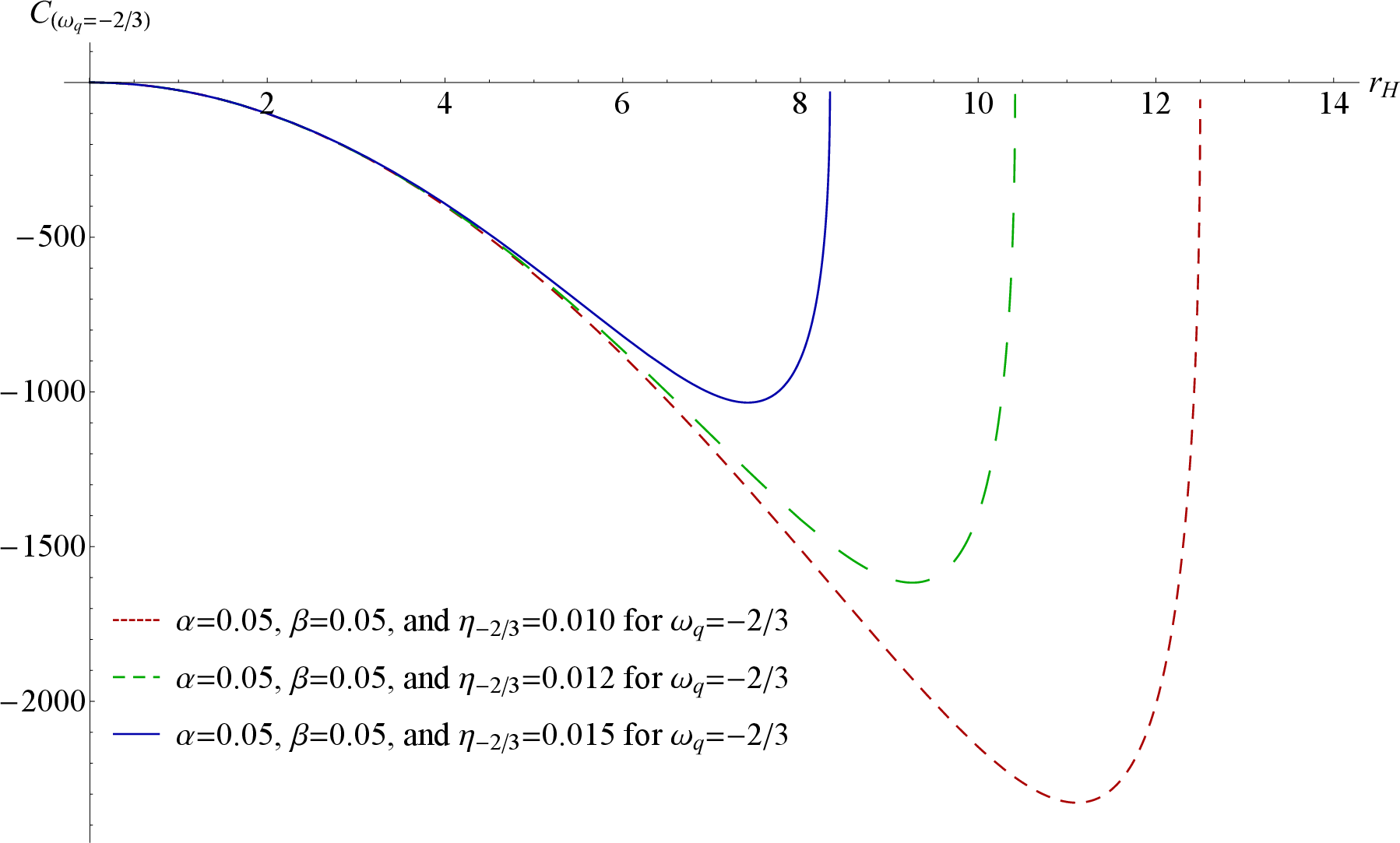}
\caption{Specific heat of the black hole against its event horizon radius for different values of $\eta_{-2/3}$.\label{Fig4}}
\end{center}
\end{figure}
One can observe from Fig.(\ref{Fig4}) that the minimum value of the specific heat becomes smaller with increasing values of $\eta_{-2/3}$.
\item For $\omega_q=-1/3$, $\frac{dM}{dr_H}$ reads
\begin{equation}\label{2.31}
\begin{split}
\frac{dM}{dr_H}=\frac{1-\eta_{-1/3}}{2}~.
\end{split}
\end{equation}
Substituting eq.(s)(\ref{2.28},\ref{2.31}) back in eq.(\ref{2.27}), we obtain the specific heat of the black hole as
\begin{equation}\label{2.32}
C_{(\omega_q=-1/3)}=\frac{\hbar}{8\pi}(1-\eta_{-1/3})\left(-\frac{1}{T^2}+\frac{\beta^2}{m_p^2}\right)~.
\end{equation}
As the term inside the parentheses remains the same after the expansion in terms of the event horizon radius of the black hole as can be seen from eq.(\ref{2.30}), we do not write them explicitly. We can again plot the specific heat of the black hole against the event horizon radius for vanishing as well as the non-vanishing case of the linear GUP parameter. We have kept the value of $\beta$ to be the same as the earlier case and have set $\eta_{-1/3}=0.01$ for the plot in Fig.(\ref{Fig5}). The value of the $\alpha$ parameter is set to $0.05$ and then zero for a side-by-side comparison. We find out from Fig.(\ref{Fig5}) that the decay rate for the specific heat is faster (as observed in the earlier case) for a non-vanishing $\alpha$ value. Unlike the previous case, the specific heat does not have a minimum and it decreases indefinitely with the increasing value of $r_H$.
\begin{figure}[ht!]
\begin{center}
\includegraphics[scale=0.28]{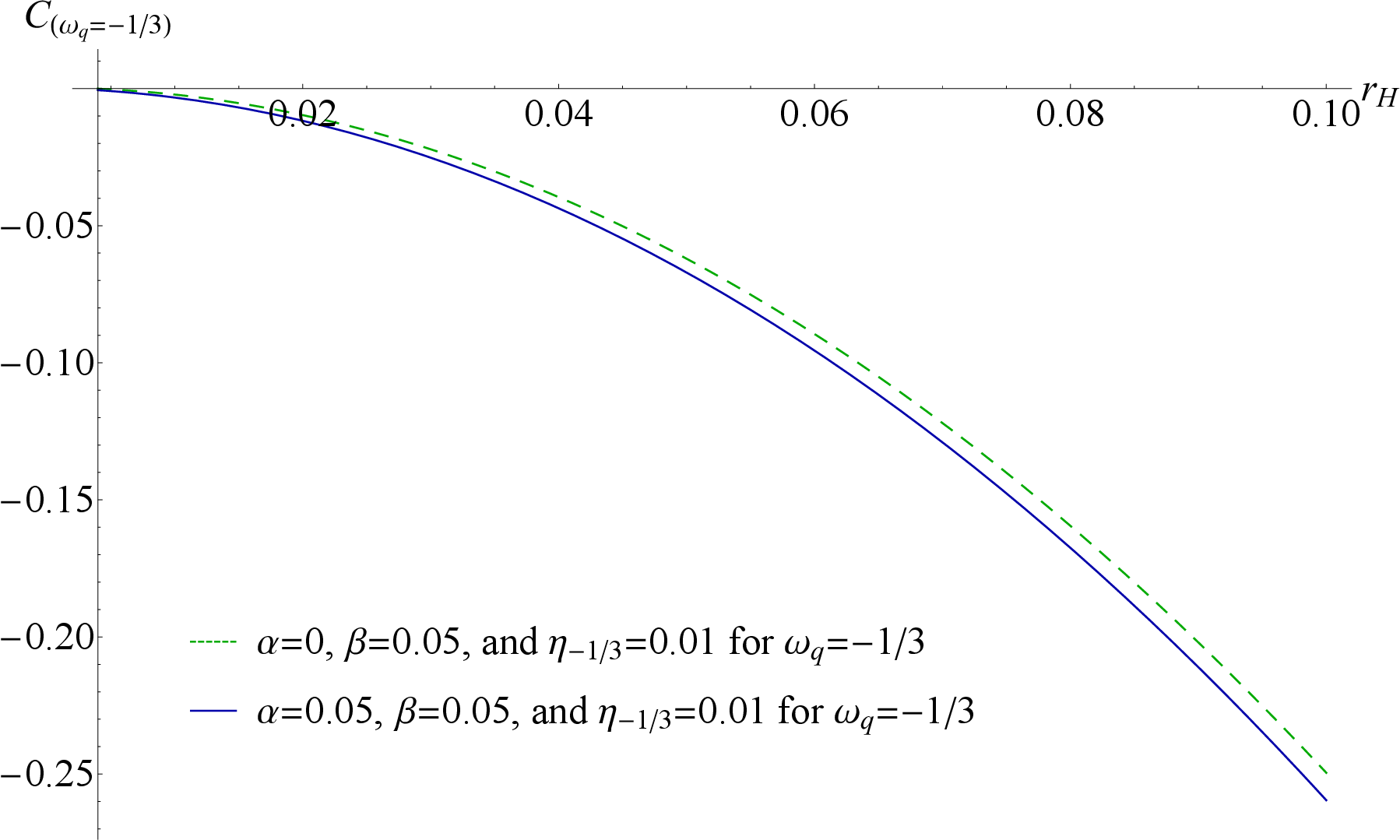}
\caption{Specific heat of the black hole against its event horizon radius for $\alpha=0$ and $\alpha\neq0$ when $\eta_{-1/3}=0.01$.\label{Fig5}}
\end{center}
\end{figure}
\item Finally for $\omega_q=-1$, we follow the same procedure and obtain the form of $\frac{dM}{dr_H}$ to be
\begin{equation}\label{2.33}
\frac{dM}{dr_H}=\frac{1-3\eta_{-1}r_H^2}{2}~.
\end{equation}
Again as before, we can obtain the form of the specific heat to be
\begin{equation}\label{2.34}
C_{(\omega_q=-1)}=\frac{\hbar}{8\pi}(1-3\eta_{-1}r_H^2)\left(-\frac{1}{T^2}+\frac{\beta^2}{m_p^2}\right)~.
\end{equation}
\end{enumerate} 
We shall now plot the specific heat of the black hole against the event horizon radius for the three cases discussed in this subsection in Fig.(\ref{Fig6}). We use $\alpha=\beta=0.05$, and $\eta_{-2/3}=\eta_{-1/3}=\eta_{-1}=0.01$ for a side-by-side comparison between the three cases. It is although important to notice that the $3\eta_{-1}r_H^2$ and $2\eta_{-2/3}r_H$ factor should remain less than unity.
\begin{figure}[ht!]
\begin{center}
\includegraphics[scale=0.28]{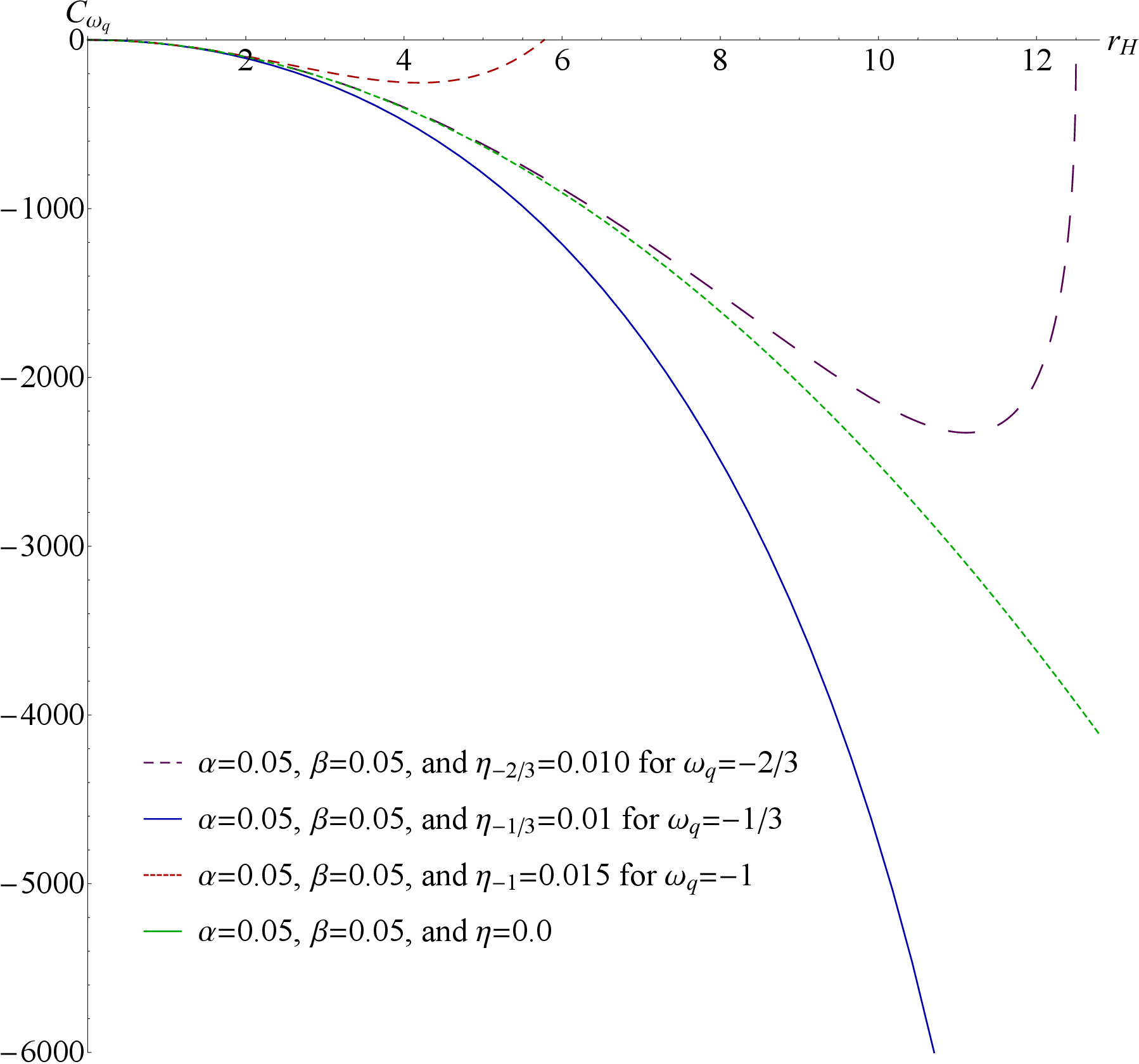}
\caption{Specific heat of the black hole against the event horizon radius for $\omega_q=-2/3,-1/3,-1$.\label{Fig6}}
\end{center}
\end{figure}
From Fig.(\ref{Fig6}), it is easy to observe that the specific heat value has the steepest fall for the $\omega_q=-1/3$ case and the slowest fall for the $\omega_q=-1$ case. It is quite easy to understand the forms of the specific heat. The specific heat for the $\omega_q=-1/3$ case has a prefactor $(1-\eta_{-1/3})$ which is not dependent on $r_H$, as a result, it falls indefinitely whereas for the $\omega_q=-1$ case the $(1-3\eta_{-1}r_H^2)$ starts to approach zero as soon as the value of the $r_H$ increases. The specific heat corresponding to the ``no-quintessence matter" case ($\eta=0$) in Fig. (\ref{Fig6}) decreases indefinitely. The important observation is that the existence of the quintessence matter around the black hole can hugely alter the behaviour of the thermodynamical quantities as can be seen from Fig.(\ref{Fig6}). For example, in the $\omega_q=-1$ case (as well as the $\omega_q=-\frac{2}{3}$ case the specific heat approach zero with increase in the event horizon radius (or the mass of the black hole) and this effect is unique to the existence of the quintessence matter and cannot be observed otherwise for a general Schwarzschild black hole.
 
\noindent In the next subsection, we shall calculate the remnant mass of the black hole for the three cases using the forms of the specific heat obtained in this subsection.
  
\subsection{Remnant Mass} \noindent From a simple physical consideration, one can argue that there exists a temperature at which the heat capacity vanishes \cite{DuttaPadhyay}. The radiation process stops at this temperature of the black hole while the black hole is left with a finite mass which is also termed as the remnant mass of the black hole. With the specific heat of the black hole obtained, we can calculate the remnant mass of the black hole for different values of the quintessential state parameter. 

\begin{enumerate}
\item For $\omega_q=-2/3$, the specific heat is obtained in eq.(\ref{2.30}). The vanishing condition of the specific heat implies 
\begin{equation}\label{2.35}
\begin{split}
\frac{1}{T^2}=&\frac{\beta^2}{m_p^2}\\\implies \frac{m_p}{\beta}=&\frac{1}{2\beta^2}\left(\alpha m_p+4\pi r_H\right)\\&-\frac{1}{2\beta^2}\sqrt{\left(\alpha m_p+4\pi r_H\right)^2-4m_P^2\beta^2}~,
\end{split}
\end{equation}
where in the last line, we have made use of the temperature equation from eq.(\ref{2.20}). Solving the above equation leads to the following relation
\begin{equation}\label{2.36}
r_H=\frac{m_p}{4\pi}(2\beta-\alpha)
\end{equation}
which is identical to the critical value of the event horizon radius obtained in eq.(\ref{2.21}).  Again using the vanishing condition of the lapse function at the event horizon radius, we obtain the remnant mass of the black hole to be
\begin{equation}\label{2.37}
M_{\text{Rem.}_{(\omega_q=-2/3)}}= \frac{m_p}{8 \pi}(2\beta - \alpha) - \frac{\eta_{-2/3}m_p^2}{32 \pi^2}(2\beta - \alpha)^2~.
\end{equation}
Following the same procedure one can obtain the remnant mass of the black hole for the other two cases which are given below.
\item For $\omega_q=-1/3$ the remnant mass of the black hole reads
\begin{equation}\label{2.38}
M_{\text{Rem.}_{(\omega_q=-1/3)}}= \frac{m_p}{8 \pi}(1-\eta_{-1/3})(2\beta - \alpha)~.
\end{equation}
\item Similarly by setting $C_{(\omega_q=-1)}=0$ from eq.(\ref{2.34}), one can obtain the remnant mass of the black hole for the $\omega_q=-1$ case as
\begin{equation}\label{2.39}
M_{\text{Rem.}_{(\omega_q=-1)}}=\frac{m_p}{8 \pi }(2\beta - \alpha)-\frac{\eta_{-1}m_p^3}{128\pi^3}(2\beta-\alpha)^3~.
\end{equation}
\end{enumerate}
It is very important to observe by comparing eq.(s)(\ref{2.22}-\ref{2.24},\ref{2.26}) with eq.(s)(\ref{2.37}-\ref{2.39}), that the critical mass of a black hole is exactly same to its remnant mass. Physically it is quite reasonable as the critical mass signifies the mass of the black hole below which the thermodynamic quantities become ill-defined or the temperature is complex-valued. On the other hand, the remnant mass indicates the mass at which the black hole has stopped radiating which means below this value, the black hole will not radiate which is equivalent to saying there is no temperature of the black hole. So it is quite intuitive and straightforward to understand that the two quantities need to be the same.
  
\subsection{Entropy}
\noindent In this subsection, we shall calculate the entropy of the black hole by using the first law of black hole thermodynamics. From the first law of black hole thermodynamics, we know the form of the entropy to be
\begin{equation}\label{2.40}
\begin{split}
S&=\int\frac{dM}{T}=\int\frac{dM}{dT}\frac{dT}{T}\\
\implies S&=\int C\frac{dT}{T}~.
\end{split}
\end{equation}
\begin{enumerate}
\item For $\omega_q=-2/3$, we obtain the form of the entropy as
\begin{equation}\label{2.41}
\begin{split}
S_{(\omega_q=-2/3)}&=\int C_{(\omega_q=-2/3)}\frac{dT}{T}\\&=\frac{\hbar}{8\pi}(1-2\eta_{-2/3}r_H)\int \left(-\frac{dT}{T^3}+\frac{\beta^2}{m_p^2}\frac{dT}{T}\right)\\
&=\frac{\hbar}{8\pi}(1-2\eta_{-2/3}r_H)\left(\frac{1}{2T^2}+\frac{\beta^2}{m_p^2}\ln\left( \frac{T}{l_p}\right)\right)
\end{split}
\end{equation}
where we have made use of eq.(\ref{2.30}) and used the form of the specific heat to calculate the entropy of the black hole.
Up to $\mathcal{O}(\alpha^4,\alpha^2\beta^2,\beta^4)$, $T$ and $\frac{1}{T^2}$ read
\begin{widetext}
\begin{align}
T&\simeq\frac{\hbar}{4\pi r_H}\left(1-\frac{\alpha m_p}{4\pi r_H}+\frac{(\alpha^2+\beta^2)m_p^2}{16 \pi^2r_H^2}-\frac{(\alpha^2+3\beta^2)\alpha m_p^3}{64\pi^3 r_H^3}+\frac{m_p^4}{256 \pi^4r_H^4}(\alpha^4+6\alpha^2\beta^2+\beta^4)\right)\label{2.42}\\
\frac{1}{T^2}&\simeq \frac{16\pi^2 r_H^2}{m_p^2}\left(1+\frac{\alpha m_p}{2\pi r_H}+\frac{(\alpha^2-2\beta^2)m_p^2}{16\pi^2r_H^2}-\frac{\beta^4 m_p^4}{256 \pi^4r_H^4}\right)~.\label{2.43}
\end{align}
\end{widetext}
The area of the black hole is given by $A=4\pi r_H^2$. Substituting eq.(s)(\ref{2.42},\ref{2.43}) in eq.(\ref{2.41}) and doing binomial expansion of the logarithmic term, one arrives at the entropy formula in terms of the area of the black hole as
\begin{widetext}
\begin{equation}\label{2.44}
\begin{split}
&S_{(\omega_q=-2/3)}\simeq \frac{A}{4l_p^2}\left(1-\frac{\alpha l_p\eta_{-2/3}}{\pi}\right)-\frac{2 l_p\eta_{-2/3}}{\sqrt{\pi}}\left(\frac{A}{4l_p^2}\right)^{\frac{3}{2}}+\frac{1}{2\sqrt{\pi}}\left(\frac{A}{4l_p^2}\right)^{\frac{1}{2}}\biggr[\alpha-\frac{l_p\eta_{-2/3}}{4\pi}(\alpha^2-2\beta^2)+\frac{l_p\eta_{-2/3}}{4\pi}\beta^2\ln(16\pi)\biggr]\\&-\frac{\alpha \beta^2}{32\pi^{\frac{3}{2}}}\left(\frac{A}{4l_p^2}\right)^{-\frac{1}{2}}+\biggr[\left(\frac{A}{4l_p^2}\right)^{-1}-\frac{2l_p\eta_{-2/3}}{\sqrt{\pi}}\left(\frac{A}{4l_p^2}\right)^{-\frac{1}{2}}\biggr]\frac{(\alpha^2+\beta^2)\beta^2}{256\pi^2}-\frac{\beta^2}{16\pi}\ln\left(\frac{A}{4 l_p^2}\right)\left(1-\frac{2 l_p\eta_{-2/3}}{\sqrt{\pi}}\left(\frac{A}{4l_p^2}\right)^{\frac{1}{2}}\right)~.
\end{split}
\end{equation}
\end{widetext}
It is important to note that the uncertainty product in eq.(\ref{1.2}) is restricted to the second order in the GUP parameters and as a result, one should not go beyond the second order while computing the entropy of the black hole. It is important to recall that the uncertainty product in the case of GUP is an approximation where higher-order contributions in the momentum uncertainty have been dropped as they are considered to be very small. As a result, it is more prudent to truncate the results of the entropy or any other thermodynamic quantity after the second-order contribution in the GUP parameter. We have kept some results up to the fourth order for some thermodynamical quantities just to inspect its behaviour at higher orders. The coefficient should change in higher order contributions of the uncertainty product if eq.(\ref{1.2}) is considered.
\begin{widetext}
\item For the $\omega_q=-1/3$ case the entropy is obtained upto $\mathcal{O}(\alpha^4,\alpha^2\beta^2,\beta^4)$ as
\begin{equation}\label{2.45}
\begin{split}
S_{(\omega_q=-1/3)}\simeq&(1-\eta_{-1/3})\biggr(\frac{\pi r_H^2}{l_p^2}+\frac{\alpha r_H}{2 l_p}+\frac{(\alpha^2-2\beta^2)}{16\pi}-\frac{\alpha \beta^2 l_p}{32\pi^2 r_H}+\frac{(\alpha^2+\beta^2)\beta^2 l_p^2}{256 \pi^3 r_H^2 }-\frac{\beta^2}{16\pi}\ln(16\pi)-\frac{\beta^2}{16\pi}\ln\left(\frac{\pi r_H^2}{l_p^2}\right)\biggr)~.
\end{split}
\end{equation}
In terms of the area of the black hole, the entropy relation above can be rewritten as
\begin{equation}\label{2.46}
\begin{split}
S_{(\omega_q=-1/3)}=&\left(1-\eta_{-1/3}\right)\left( \frac{A}{4l_p^2}+\frac{\alpha}{2\sqrt{\pi}}\left( \frac{A}{4l_p^2}\right)^{\frac{1}{2}}-\frac{\beta^2}{16\pi}\ln\left( \frac{A}{4l_p^2}\right)-\frac{\alpha\beta^2}{8\pi^{\frac{3}{2}}}\left( \frac{A}{4l_p^2}\right)^{-\frac{1}{2}}+\frac{(\alpha^2+\beta^2)\beta^2}{256\pi^2}\left( \frac{A}{4l_p^2}\right)^{-1}+K_0\right)~,
\end{split}
\end{equation}
where 
\begin{equation}\label{2.47}
K_0\equiv\frac{1-\eta_{-1/3}}{16\pi}\left(\alpha^2-2\beta^2-\beta^2\ln(16\pi)\right)~.
\end{equation}
\item  For $\omega_q=-1$, the entropy for the black hole in terms of its area reads
\begin{equation}\label{2.48}
\begin{split}
S_{(\omega_q=-1)}=&\frac{A}{4l_p^2}\left[1-\frac{3l_p^2\eta_{-1}}{16\pi^2}\left(\alpha^2-(2+\ln(16\pi))\beta^2\right)\right]-\frac{3l_p^2\eta_{-1}}{4\pi}\left(\frac{A}{4l_p^2}\right)^2-\frac{3l_p^2\eta_{-1}\alpha}{2\pi^{\frac{3}{2}}}\left(\frac{A}{4l_p^2}\right)^{\frac{3}{2}}+\frac{\alpha}{2\sqrt{\pi}}\left(\frac{A}{4l_p^2}\right)^{\frac{1}{2}}\Bigr(1\\&+\frac{3 l_p^2\eta_{-1}\beta^2}{4\pi^2}\Bigr)-\frac{\beta^2}{16 \pi}\left(1-\frac{3 l_p^2\eta_{-1}}{\pi}\left(\frac{A}{4l_p^2}\right)\right)\ln\left(\frac{A}{4l_p^2}\right)-\frac{\alpha \beta^2}{8\pi^{\frac{3}{2}}}\left(\frac{A}{4l_p^2}\right)^{-\frac{1}{2}}+\frac{\left(\alpha^2+\beta^2\right)\beta^2}{256\pi^2}\left(\frac{A}{4l_p^2}\right)^{-1}+K_1~,
\end{split}
\end{equation}
where 
\begin{equation}\label{2.49}
K_1=-\frac{3l_p^2\eta_{-1}\beta^2}{256\pi^3}\beta^2(\alpha^2+\beta^2)+\frac{1}{16\pi}\left(\alpha^2-(2+\ln(16\pi))\beta^2\right)~.
\end{equation}
\end{widetext}
In eq.(\ref{2.46},\ref{2.48}) $K_0$ and $K_1$ terms do not depend on the area of the black hole as a result these terms can be neglected from the expressions of the entropy terms. Here, we have kept these terms for the sake of the completeness of the results.
From eq.(s)(\ref{2.44},\ref{2.46},\ref{2.48}), we observe that the entropy of the black hole carries logarithmic as well as inverse order correction terms in the area of the black hole. This is solely a consequence of the generalized uncertainty principle framework and all such correction terms vanish in the $\alpha,\beta\rightarrow0$ limit.
 \end{enumerate}
 
 \subsection{Energy Density}
 \noindent With the form of the entropy in hand corresponding to the three constant values of the quintessential state parameter, we are now in a position to calculate the energy density of the black hole surrounded by quintessence matter in the LQGUP framework for the three separate cases.
In the $\alpha,\beta\rightarrow0$, limit the entropy of the black hole reads
\begin{equation}\label{3.E.1}
\begin{split}
S_\text{HUP}=\frac{\pi r_H^2}{\hbar}~.
\end{split}
\end{equation}
Using the above equation, we can rewrite eq.(\ref{2.5}), in terms of the entropy of the black hole as
\begin{equation}\label{3.E.2}
\begin{split}
\rho_q&=-\frac{3\eta_{\omega_{q}}\omega_{q}}{2r_H^{3(\omega_q+1)}}\\
&=-\frac{3\eta_{\omega_{q}}\omega_{q}}{2}\left(\frac{\pi}{\hbar S_{\text{HUP}}}\right)^{\frac{3(\omega_q+1)}{2}}~.
\end{split}
\end{equation}
In the LQGUP framework, we assume that the entropy of the black hole can still be represented as
$S_\text{GUP}=S_{\omega_q}=\frac{\pi r_{\text{GUP}}^2}{\hbar}$ where $r_{\text{GUP}}$ denotes the effective event horizon radius such that it encapsulates all of the extra contributions to the Bekenstein-Hawking entropy of the black hole for considering the GUP framework. Hence, the modified energy density in the LQGUP framework takes the form
\begin{equation}\label{3.E.3}
\rho_q=-\frac{3\eta_{\omega_{q}}\omega_{q}}{2}\left(\frac{\pi}{\hbar S_{\omega_q}}\right)^{\frac{3(\omega_q+1)}{2}}~.
\end{equation}
\begin{enumerate}
\item For the $\omega_q=-2/3$ case, eq.(\ref{3.E.3}) can be recast in the following form
\begin{equation}\label{3.E.4}
\rho_q=\eta_{-2/3}\sqrt{\frac{\hbar S_{\omega_q=-2/3}}{\pi}}~.
\end{equation}
For the unit choices made in our current analysis $\hbar=l_p^2$, we can therefore obtain the form of the energy density up to $\mathcal{O}(\eta\alpha^2,\eta\beta^2)$ by using the form of the entropy in eq.(\ref{2.44}) as
\begin{equation}\label{3.E.5}
\rho_q\simeq r_H\eta_{-2/3}\left[1+\frac{\alpha l_p}{4\pi r_H}-\frac{l_p^2\left[\alpha^2+\beta^2\ln\left[\frac{\pi r_H^2}{l_p^2}\right]\right]}{32\pi^2r_H^2}\right]
\end{equation}
where we have dropped all $\mathcal{O}(\eta^2)$ contributions. We shall now compare the energy density of the black hole in the presence and absence of the linear GUP parameter in Fig.(\ref{Fig7}).
\begin{figure}[ht!]
\begin{center}
\includegraphics[scale=0.28]{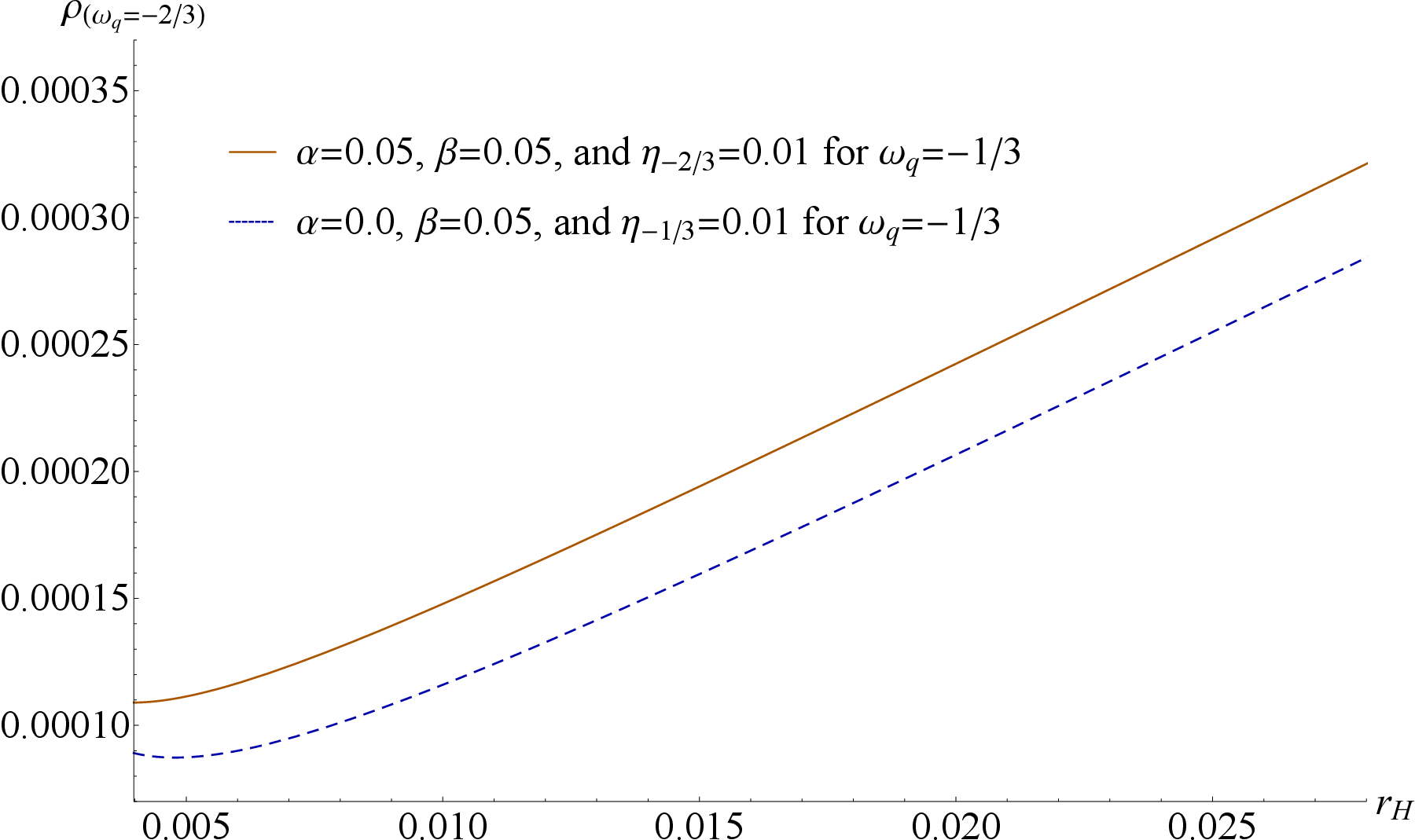}
\caption{Energy density vs $r_H$ plot for $\omega_q=-2/3$ when $\alpha=0.05$ and $\alpha=0$.\label{Fig7}.}
\end{center}
\end{figure}
It is straightforward to observe from Fig.(\ref{Fig7}) that the energy density is higher when the linear GUP parameter is present than the case when it is absent. 
\item For $\omega_q=-\frac{1}{3}$, eq.(\ref{3.E.3}) can be recast as
\begin{equation}\label{3.E.6}
\begin{split}
\rho_q=\frac{\pi\eta_{-1/3}}{2\hbar S_{\omega_q=-1/3}}~.
\end{split}
\end{equation}
Using eq.(\ref{2.45}), we can recast the expression of the above energy density as
\begin{equation}\label{3.E.7}
\begin{split}
\rho_q\simeq\frac{\sqrt{\pi}\eta_{-1/3}}{2l_pr_H}\left[1-\frac{\alpha l_p}{2\pi r_H}+\frac{l_p^2\left[\alpha^2+\frac{\beta^2}{4}\ln\left[\frac{\pi r_H^2}{l_p^2}\right]\right]}{4\pi^2r_H^2}\right]~.
\end{split}
\end{equation}
Using the above approximate result we shall compare again between the QGUP (quadratic GUP with $\alpha=0$) and LQGUP case in Fig.(\ref{Fig8}).
\begin{figure}[ht!]
\begin{center}
\includegraphics[scale=0.28]{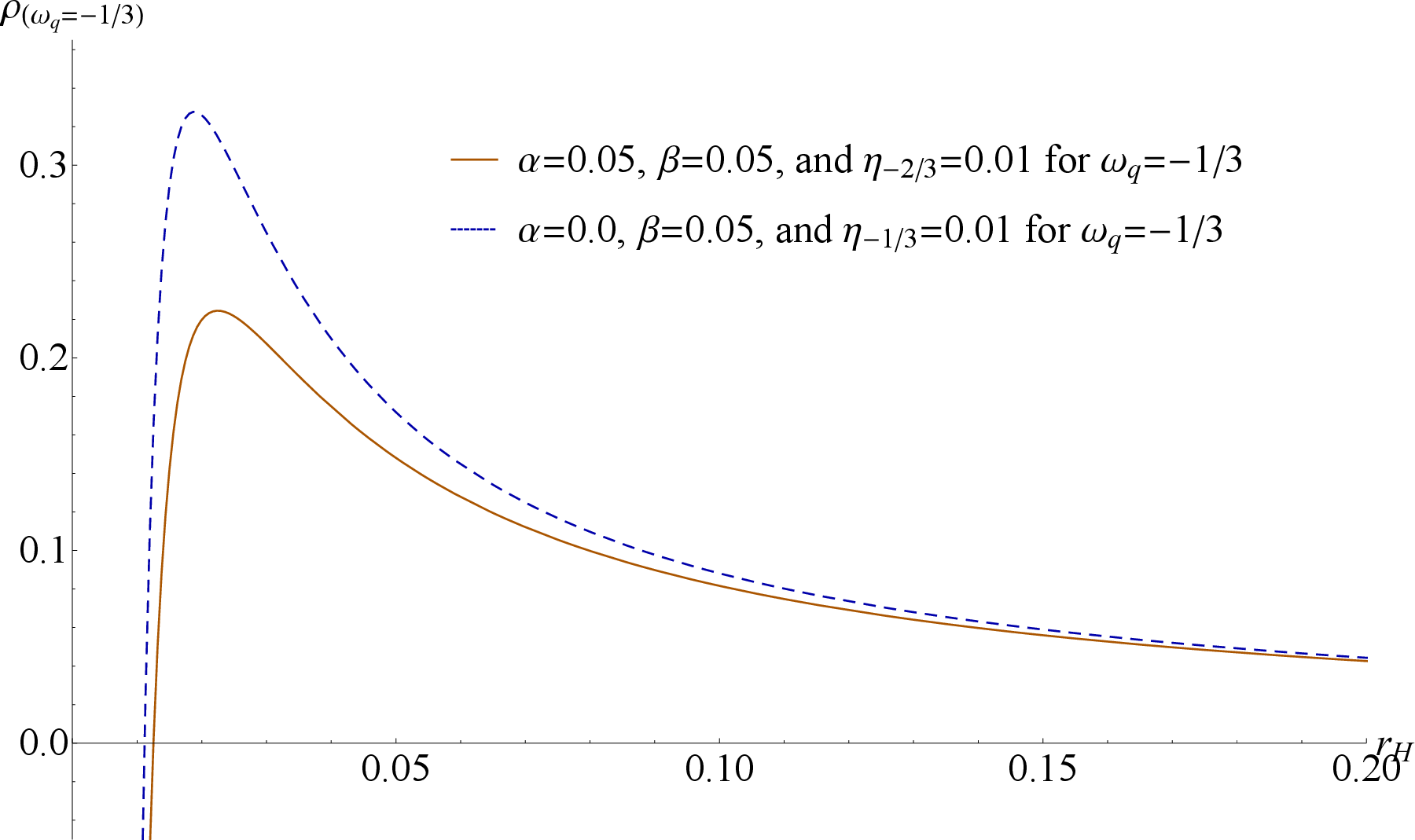}
\caption{$\rho_q$ vs $r_H$ plot $\omega_q=-1/3$. We compare the QGUP case with the LQGUP case and observe the effect of the linear GUP parameter\label{Fig8}.}
\end{center}
\end{figure}
We observe from Fig.(\ref{Fig8}) that, unlike the $\omega_q=-2/3$ case, the energy density shows a slower growth and decay rate with increasing $r_H$ when $\alpha=0.05$ than the $\alpha=0$ case.
We shall now plot the energy densities for the above two cases against the event horizon radius of the black hole in Fig.(\ref{Fig9}).
\begin{figure}[ht!]
\begin{center}
\includegraphics[scale=0.28]{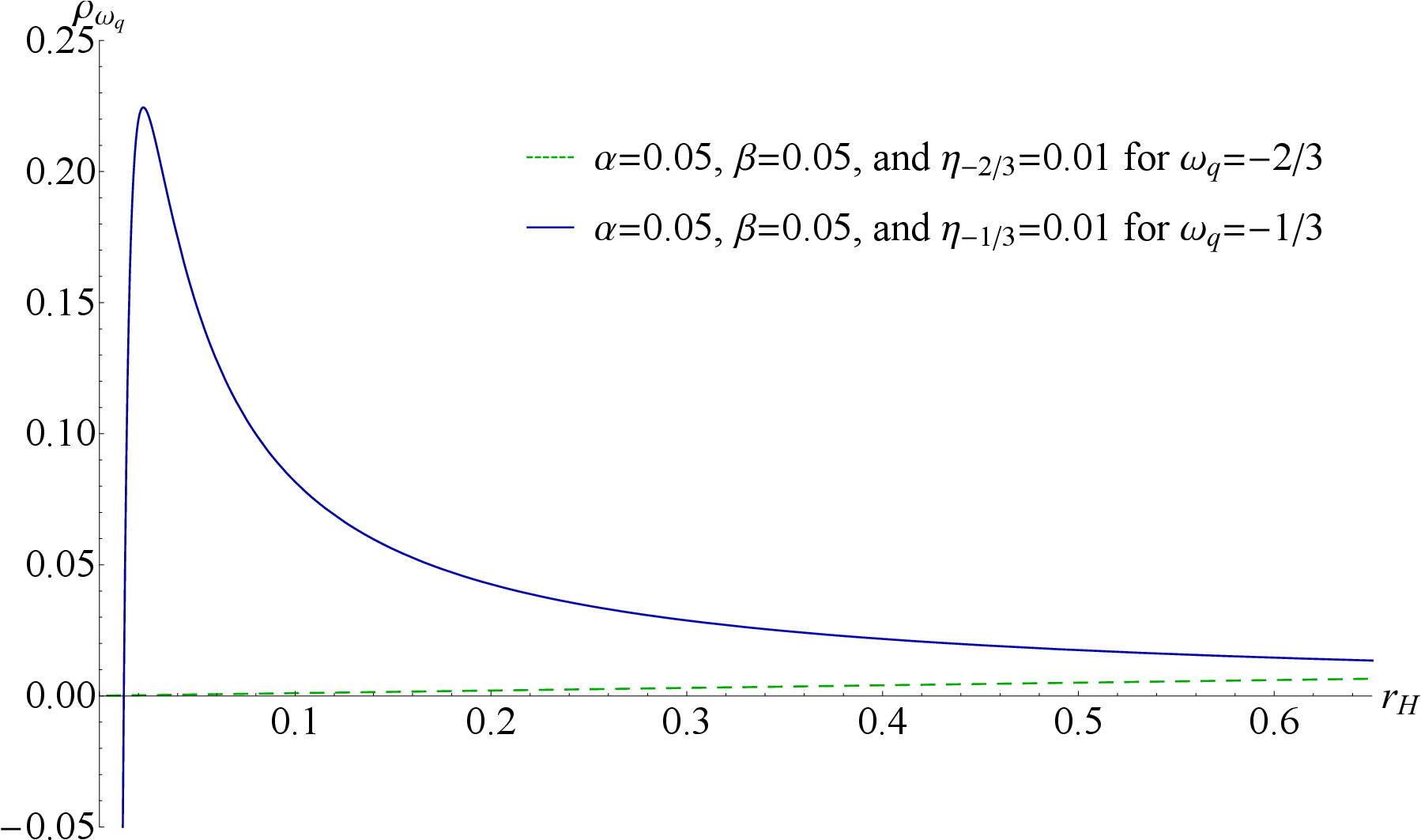}
\caption{Energy density is plotted against the event horizon radius for $\omega_q=-2/3$ and $\omega_q=-1/3$\label{Fig9}.}
\end{center}
\end{figure}
From Fig.(\ref{Fig9}), we observe that the energy density of the black hole first grows rapidly, then decays for the $\omega_q=-1/3$ case when plotted against $r_H$ where $\rho_q$ grows gradually for the $\omega_q=-2/3$. The reason for this behaviour primarily is the overall $r_H$ term being present in the $\rho_q$ term when $\omega_q=-2/3$. The parameters are again kept the same as used to plot Fig.(\ref{Fig6}).
\item Finally for the $\omega_q=-1$ case,  eq.(\ref{3.E.3}) can be recast as
\begin{equation}\label{3.E.8}
\rho_q=\frac{3\eta_{-1}}{2}~.
\end{equation}
The above equation implies that for $\omega_q=-1$ case the energy density is independent of the event horizon radius of the black hole and is a constant dependent only upon the quintessential normalization factor $\eta_{-1}$. The pressure of the black hole can be directly obtained using eq.(\ref{2.6}). Hence, one can derive the pressure for each case discussed above by just multiplying the corresponding value of $\omega_q$ with the energy density obtained in this subsection. 
\end{enumerate}

 \section{Energy output as a function of time }\label{S4}
\noindent While a black hole radiates, the mass of the black hole decreases gradually. With the decreasing of its mass, the temperature of the black hole increases. One can assume that this energy loss is dominated via photons and as a result the standard Stefan-Boltzmann law can be applied to estimate the energy radiated as a function of time
 \begin{equation}\label{4.1}
 	\frac{dM}{dt} = -\sigma A {T_H}^4~,
 \end{equation}
 where $\sigma$ is the Stefan-Boltzmann constant. The above relation can be reinterpreted as
\begin{equation}\label{4.2}
\frac{dM}{dr_H}\frac{dr_H}{dt}=-\sigma (4\pi r_H^2)T_H^4
\end{equation}
where $A=4\pi r_H^2$ denotes the area of the black hole. For a general Schwarzschild black hole in the $\eta_{\omega_q},\alpha,\beta\rightarrow0$ limit, one can recast eq.(\ref{4.2}) as
\begin{equation}\label{4.3}
\frac{1}{2}\frac{dr_{H_{0}}}{dt}=-\sigma (4\pi r_{H_0}^2)T_{H_0}^4~,
\end{equation}
where $r_{H_0}=2M$ denotes the Schwarzschild radius and $T_{H_0}=\frac{\hbar}{4\pi r_{H_0}}$ denotes the Hawking temperature for a Schwarschild black hole. Setting $r_{H_0}=2 \sqrt{\hbar} x=2m_p x $ one can recast eq.(\ref{4.3}) as
\begin{equation}\label{4.4}
\begin{split}
\frac{dx}{dt}=-\frac{\sigma m_p^5}{256\pi^3}\frac{1}{x^2}~.
\end{split}
\end{equation}
Defining the characteristic time as  $t_{\text{ch}}=\frac{256\pi^3}{\hbar^\frac{5}{2}\sigma}$ \cite{DuttaPadhyay5}, we can recast eq.(\ref{4.4}) as
\begin{equation}\label{4.5}
\frac{dx}{dt}=-\frac{1}{t_{\text{ch}}x^2}~. 
\end{equation}
The solution of the above equation yields the relation
\begin{equation}\label{4.6}
\frac{x^3}{3}+A=-\frac{t}{t_{\text{ch}}}~,
\end{equation}
where $A$ is an undetermined constant. Now at $t=0$ if  $x(0)=x_i$,  then the solution of  eq.(\ref{4.5}) in eq.(\ref{4.6}) yields the mass-time relation to be
 \begin{equation}\label{4.7}
 x = \left(x_i^3 - \frac{3t}{t_{\text{ch}}}\right)^\frac{1}{3}~.
 \end{equation}
At $t=0$, $x_i=\frac{M(0)}{m_p}$, and as a result $m_p x_i$ denotes the initial mass $M(0)$ of the black hole. The rate at which energy is radiated as a 
 function of time now takes the form 
 \begin{equation}\label{4.8}
 	\frac{dx}{dt} = -\frac{1}{t_{ch}\left(x_i^3 - \frac{3t}{t_{ch}}\right)^\frac{2}{3}}
 \end{equation}
 which is obtained by substituting eq.(\ref{4.7}) in the right hand side of eq.(\ref{4.5}).  If the black hole evaporates completely, then $x=0$. The evaporation time is therefore given by
\begin{equation}\label{4.9}
t_{\mathcal{E}}=\frac{t_{\text{ch}}}{3}\left(\frac{M(0)}{m_p}\right)^3~.
\end{equation}
Our primary aim is to calculate the mass-time relation and the evaporation time of the Schwarzschild black hole surrounded by the quintessence energy-matter in the LQGUP framework for the three fixed values of the quintessential state parameter.
 %In terms of Schwarzschild black hole horizon radius $r_H$ with the horizon area  $A=4\pi r_{H}^2 $ and for $\omega_{q}= -2/3 $, above equation implies
%R \begin{eqnarray}
 %	\left(\frac{1-2\alpha r_H}{2}\right)\frac{dr_H}{dt} =
% 	-\sigma \left(4\pi r_{H}^2\right) {T_H}^4
% 	\label{Stefan-Boltzmann law_omega_q=-2/3}
 %\end{eqnarray}
 %In a similar manner by using eq.(\ref{Stefan-Boltzmann law defination}), for $\omega_{q}= -1/3 $ and for $\omega_{q}= -1$, we have the following equations 
 %\begin{eqnarray}
 	%\left(\frac{1-\alpha}{2}\right)\frac{dr_H}{dt} =
% 	-\sigma \left(4\pi r_{H}^2\right) {T_H}^4~,\\
 %	\label{Stefan-Boltzmann law_omega_q=-1/3}
% 	\left(\frac{1}{2+24\alpha M^2}\right)\frac{dr_H}{dt} =
 %	-\sigma \left(4\pi r_{H}^2\right) {T_H}^4~,
 	%\label{Stefan-Boltzmann law_omega_q=-1}
 %\end{eqnarray}
 %\noindent respectively.

 \subsection{Energy output as a function of time for $\omega_q=-2/3$}
\begin{figure}[ht!]
\begin{center}
\includegraphics[scale=0.28]{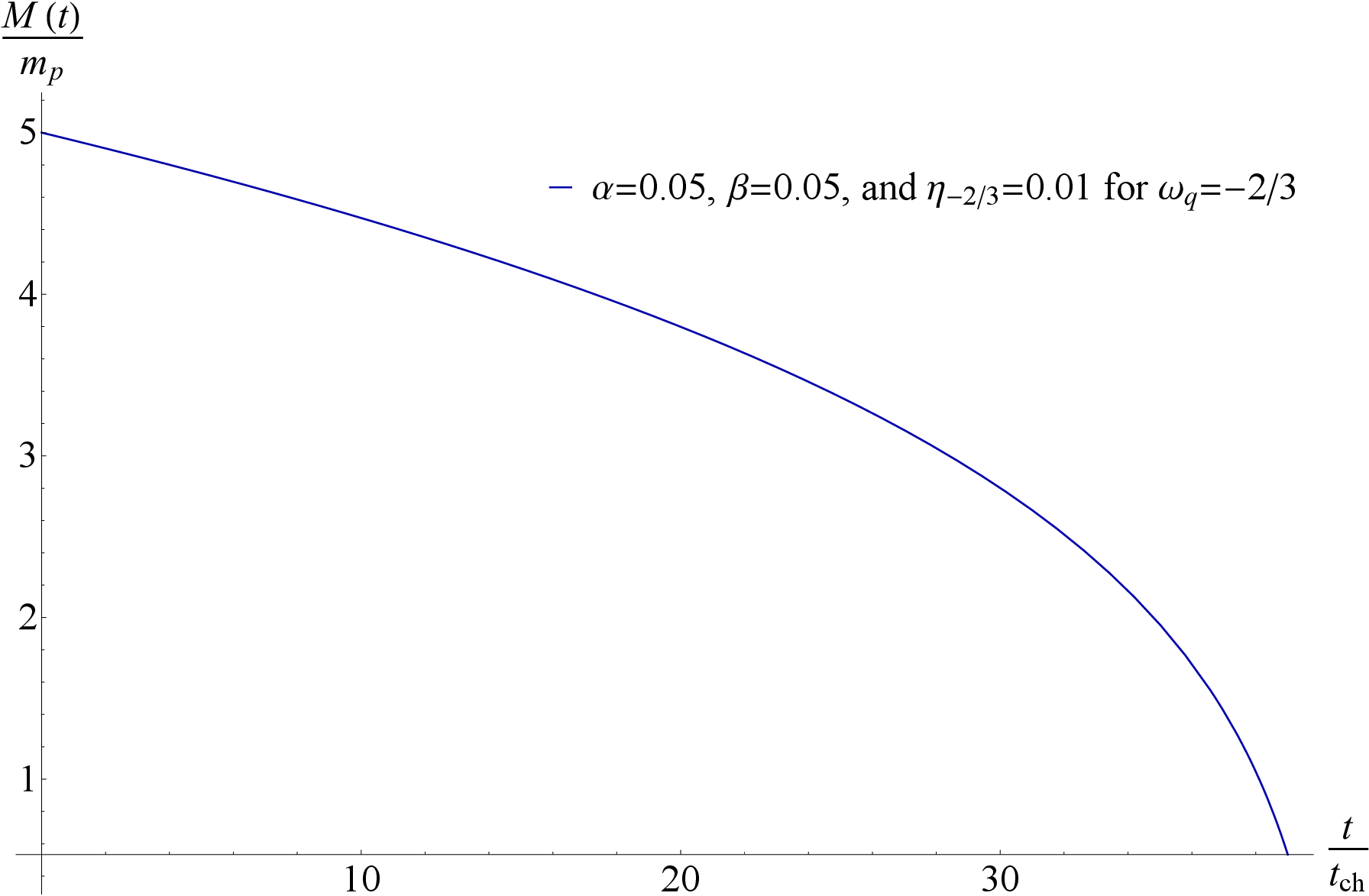}
\caption{$x(t)=\frac{M(t)}{m_p}$ is plotted against the dimensionless time $\frac{t}{t_{\text{ch}}}$ for $\alpha=\beta=0.05$ and $\eta_{-2/3}=0.01$\label{Fig10}.}
\end{center}
\end{figure}

 For $\omega_q=-\frac{2}{3}$, the evaporation equation in eq.(\ref{4.2}) reads
 \begin{equation}\label{4.10}
 \frac{1-2\eta_{-2/3}r_H}{2}\frac{dr_H}{dt}=-4\pi\sigma r_H^2 T_H^4~.
 \end{equation}
 Again using a new variable $x=\frac{r_H}{2\sqrt{\hbar}}$, up to $\mathcal{O}(\alpha^4,\alpha^2\beta^2,\beta^4)$, the above equation can be recast as
 \begin{equation}\label{4.11}
 \left(C_2'x^2+C_3 x^3+C_1 x+C_0+\frac{C_{-1}}{x}+\frac{C_{-2}}{x^2}\right) \frac{dx}{dt}\simeq-\frac{1}{t_{\text{ch}}}
 \end{equation}
 where the constant factors are given by
 \begin{equation}\label{4.12}
\begin{split}
C_3&=-4\eta_{-2/3}\sqrt{\hbar}~,\\
C_2'&=1+C_2=1-\frac{2\eta_{-2/3}\alpha\sqrt{\hbar}}{\pi}~,\\
C_1&=\frac{\alpha}{2\pi}-\frac{3\alpha^2-2\beta^2}{8\pi^2}\eta_{-2/3}\sqrt{\hbar}~,\\
C_0&=\frac{3\alpha^2-2\beta^2}{32\pi^2}-\frac{\alpha(\alpha^2-2\beta^2)}{32\pi^3}\eta_{-2/3}\sqrt{\hbar}~,\\
C_{-1}&=\frac{\alpha(\alpha^2-2\beta^2)}{128\pi^3}-\frac{(\alpha^4-4\alpha^2\beta^2+2\beta^4)}{1024\pi^4}\eta_{-2/3}\sqrt{\hbar}~,\\
C_{-2}&=\frac{(\alpha^4-4\alpha^2\beta^2+2\beta^4)}{4096\pi^4}~.
\end{split} 
\end{equation}
Integrating eq.(\ref{4.11}), we obtain 
\begin{widetext}
\begin{equation}\label{4.13}
\begin{split}
&(1+C_2)\frac{x^3}{3}+\frac{C_3x^4}{4}+\frac{C_1x^2}{2}+C_0 x+C_{-1}\ln x-\frac{C_{-2}}{x}+A_0=-\frac{t}{t_{\text{ch}}}~.
\end{split}
\end{equation}
\end{widetext}
The constant $A_0$ can be fixed by setting $t=0$ in the above equation, which gives
\begin{equation}\label{4.14}
\begin{split}
A_0=\frac{C_{-2}}{x_i}-\frac{C_2'x_i^3}{3}-\frac{C_3x_i^4}{4}-\frac{C_1x_i^2}{2}-C_0 x_i-C_{-1}\ln x_i~.
\end{split}
\end{equation}
The zeroth order solution for $x$ from eq.(\ref{4.13}) in the limit $\eta_{-2/3},\alpha,\beta\rightarrow0$ goes to eq.(\ref{4.7}). We consider $\eta_{-2/3}$ to be a small quantity and while solving for $x$, we will get rid of any $\mathcal{O}(\eta^2)$ contributions. We shall also drop any contributions higher than $\mathcal{O}(\alpha^2,\beta^2,\alpha\beta)$ for the solution $x$ as the uncertainty relation captures up to quadratic order contributions in the $\alpha,\beta$ parameters as can be seen from eq.(\ref{1.2}). We shall now solve eq.(\ref{4.13}) and obtain $x$ as a function of time upto $\mathcal{O}(\alpha^2,\beta^2,\eta\alpha^2,\eta\beta^2)$. Eq.(\ref{4.13}) can now be recast as
\begin{widetext}
\begin{equation}\label{4.15}
\begin{split}
x^3=&x_i^3-\frac{3t}{t_{\text{ch}}}+3\eta_{-2/3}\sqrt{\hbar}(x^4-x_i^4)-3\alpha\biggr(\frac{x^2-x_i^2}{4\pi}-\frac{2\eta_{-2/3}\sqrt{\hbar}(x^3-x_i^3)}{3\pi}\biggr)-\frac{9\alpha^2-6\beta^2}{32\pi^2}\bigr(x-x_i-2\eta_{-2/3}\sqrt{\hbar}(x^2-x_i^2)\bigr)~.
\end{split}
\end{equation}
\end{widetext}
We shall solve the above equation perturbatively. We start by considering that $x$ can be decomposed as
\begin{equation}\label{4.16}
x=x_0+x_\eta+x_1+x_2
\end{equation}
where the subscript of $x$ denotes the order of the term ($x_\eta$ denotes the zeroth order contribution multiplied by $\eta$). In the $\eta_{-2/3},\alpha,\beta\rightarrow0$ limit, eq.(\ref{4.15}) can be recast as
\begin{equation}\label{4.17}
\begin{split}
x_0^3&=x_i^3-\frac{3t}{t_{\text{ch}}}\\
\implies x_0&=\left(x_i^3-\frac{3t}{t_{\text{ch}}}\right)^{\frac{1}{3}}
\end{split}
\end{equation}
which is the solution of $x$ obtained in eq.(\ref{4.7}).  We now move towards obtaining the next order solution or the $x_\eta$ contribution from eq.(\ref{4.16}). In the $\alpha,\beta\rightarrow0$ limit, eq.(\ref{4.15}) reduces to
\begin{equation}\label{4.18}
\begin{split}
x=x_0\left(1+\frac{3\eta_{-2/3}\sqrt{\hbar}}{x_0^3}(x^4-x_i^4)\right)^{\frac{1}{3}}~.
\end{split}
\end{equation}
In this $\alpha,\beta\rightarrow0$ limit, eq.(\ref{4.16}) reduces to 
\begin{equation}\label{4.19}
x=x_0+x_\eta~.
\end{equation}
Substituting eq.(\ref{4.19}) in eq.(\ref{4.18}), we obtain upto $\mathcal{O}(\eta)$
\begin{equation}\label{4.20}
\begin{split}
x_0+x_\eta&\simeq x_0\left(1+\frac{\eta_{-2/3}\sqrt{\hbar}}{x_0^3}(x_0^4-x_i^4)\right)\\
\implies x_\eta &=\frac{\eta_{-2/3}\sqrt{\hbar}}{x_0^2}(x_0^4-x_i^4)~.
\end{split}
\end{equation}
Going to the next order ($\mathcal{O}(\alpha)$), the solution for $x_1$ can be obtained following the same procedure to be
\begin{widetext}
\begin{equation}\label{4.21}
\begin{split}
x_1=&-\frac{\alpha}{4\pi x_0^2}(x_0^2-x_i^2)-\frac{\alpha\eta_{-2/3}\alpha\sqrt{\hbar}}{6\pi x_0^5}(2x_0^6-3x_i^2x_0^4+4x_i^3x_0^3-3x_i^6)~.
\end{split}
\end{equation}
Finally, the second order solution to $x$ can be obtained by using eq.(s)(\ref{4.17},\ref{4.18},\ref{4.20},\ref{4.21}) in eq.(\ref{4.16}). The result is
\begin{equation}\label{4.22}
\begin{split}
x_2&=\frac{\alpha^2}{16\pi^2 x_0^5}(x_0^4-x_i^4)-\frac{3\alpha_2-2\beta^2}{32\pi^2x_0^2}(x_0-x_i)-\frac{\eta_{-2/3}\sqrt{\hbar}}{96\pi^2 x_0^8}(3x_0^8(3\alpha^2-2\beta^2)-6x_ix_0^7(3\alpha^2
-2\beta^2)+2x_i^2x_0^6(13\alpha^2-6\beta^2)
\\&+3x_i^4x_0^4(3\alpha_2-2\beta^2)-2x_i^5x_0^3
(25\alpha^2-6\beta^2)+30x_i^8\alpha^2)~.
\end{split}
\end{equation}
\end{widetext}
Hence, we have obtained the solution of $x$ as a function of time up to second order in the dimensionless GUP parameters\footnote{It is important to note that the time dependence creeps in the solution explicitly via the $x_0$ term.}. We shall now plot the solution of $x$ (obtained up to second order in the GUP parameter) against time which signifies the change of the mass of the black hole with time in Fig.(\ref{Fig10}). For simplicity, we consider the initial mass of the black hole to be $M_i=M(0)=5m_p$. 
We observe from Fig.(\ref{Fig10}) that the mass decreases gradually with time then the decay process speeds up which implies a higher evaporation rate with increasing time. 
\begin{figure}[ht!]
\begin{center}
\includegraphics[scale=0.28]{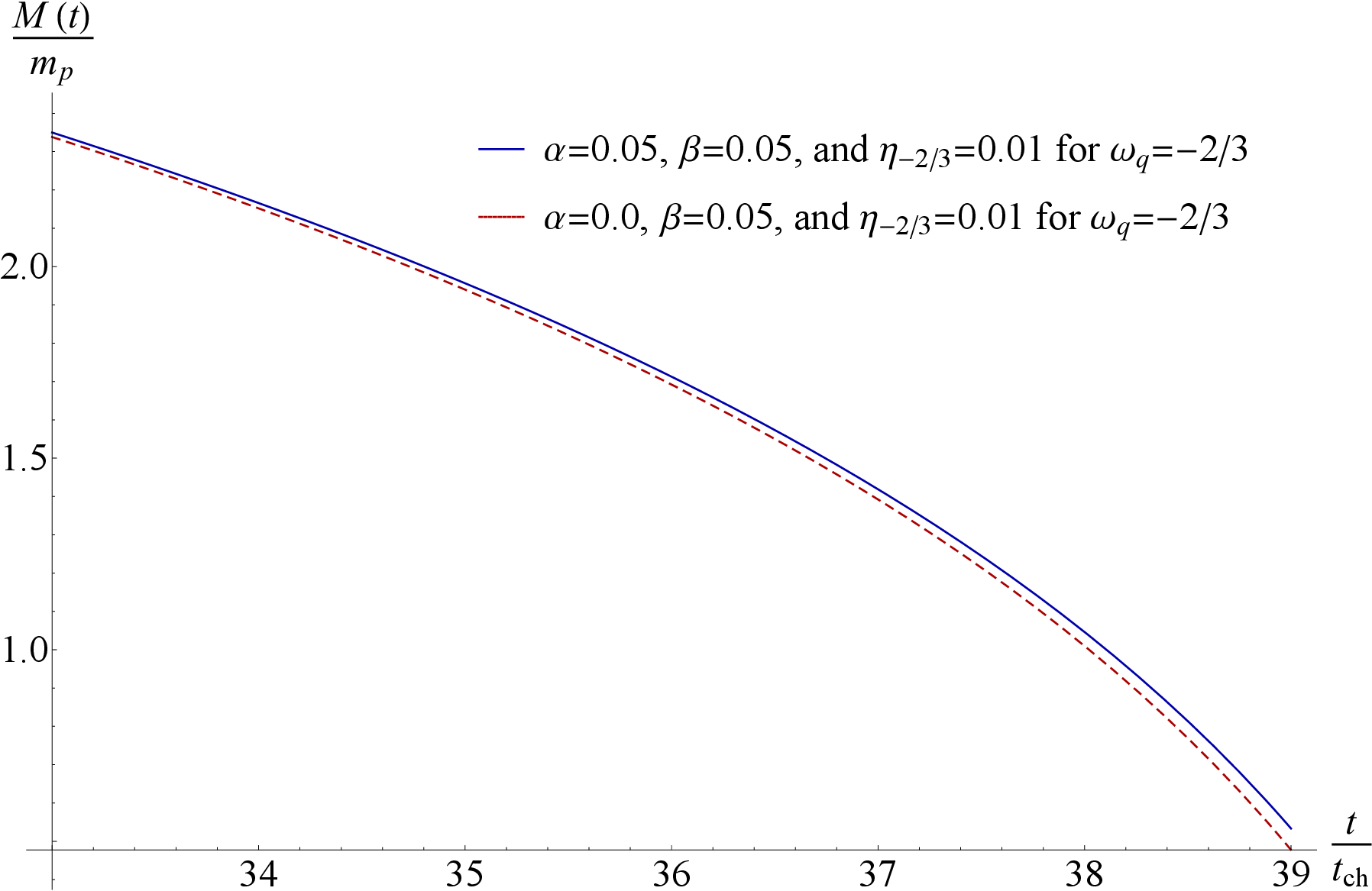}
\caption{$x(t)=\frac{M(t)}{m_p}$ is plotted against the dimensionless time $\frac{t}{t_{\text{ch}}}$ for the QGUP and the LQGUP cases\label{Fig11}.}
\end{center}
\end{figure}
Next in Fig.(\ref{Fig11}), we have plotted $x$ against $\frac{t}{t_{\text{ch}}}$ for the QGUP ($\alpha=0$) as well as the LQGUP case. We observe that the mass decays faster for the QGUP case than the LQGUP case which indicates a quicker evaporation of the black hole in the QGUP case.

\noindent Our next aim is to calculate the time at which the evaporation process stops completely. This time is also known as the evaporation time. From the form of the remnant mass in eq.(\ref{2.37}) for the $\omega_q=-2/3$ case, we observe that the black hole does evaporate completely in the $\alpha,\beta\rightarrow0$ limit. This indicates that the final value of $x$ will be zero. In the LQGUP framework, the remnant mass is of the order of the Planck mass multiplied by GUP parameters. When the evaporation process stops, the value of $x$ will be
\begin{equation}\label{4.23}
\begin{split}
x(t_{\mathcal{E}})&=\frac{M_{\text{Rem}}}{m_p}=\frac{2\beta-\alpha}{8\pi}-\frac{\eta_{-2/3}m_p}{32\pi^2}(2\beta-\alpha)^2\\
\implies x_f&=x_{f_1}-2\eta_{-2/3}m_px_{f_1}^2
\end{split}
\end{equation}
 where $t_{\mathcal{E}}$ denotes the time when the black hole stops evaporating and we have defined $x_f\equiv x(t_\mathcal{E})$, and $x_{f_1}=\frac{2\beta-\alpha}{8\pi}$. The evaporation time is obtained by integrating eq.(\ref{4.11}) from $x_i$ to $x_f$. The evaporation time reads
\begin{widetext}
\begin{equation}\label{4.24}
\begin{split}
\frac{t_{\mathcal{E}}}{t_{\text{ch}}}=\frac{x_i^3-x_f^3}{3}(1+C_2)+\frac{C_3(x_i^4-x_f^4)}{4}+\frac{C_1(x_i^2-x_f^2)}{2}+C_0(x_i-x_f)+C_{-1}\ln\left(\frac{x_i}{x_f}\right)-C_{-2}\left(\frac{1}{x_i}-\frac{1}{x_f}\right)~.
\end{split} 
\end{equation}
The above expression up to $\mathcal{O}(\alpha^4,\alpha^2\beta^2,\beta^4)$ reads
\begin{equation}\label{4.25}
\begin{split}
\frac{t_{\mathcal{E}}}{t_{\text{ch}}}&=\frac{x_i^3}{3}(1+C_2)+\frac{C_3x_i^4}{4}+\frac{C_1(x_i^2)}{2}+C_0x_i+C_{-1}\ln x_i-\frac{C_{-2}}{x_i}-\frac{x_{f_1}^3}{3}+3\eta_{-2/3}l_px_{f_1}^4+\frac{5\eta_{-2/3}\alpha l_p}{3\pi}x_{f_1}^3-\frac{\alpha x_{f_1}^2}{4\pi}\\&+\frac{3\alpha^2-2\beta^2}{8\pi^2}\eta_{-2/3}l_px_{f_1}-\frac{3\alpha^2-2\beta^2}{32\pi^2}x_{f_1}^2+\frac{3\alpha(\alpha^2-2\beta^2)}{64\pi^3}\eta_{-2/3}l_px_{f_1}+\frac{\alpha(\alpha^2-2\beta^2)}{128\pi^3}\ln x_{f_1}+\frac{\alpha^4-4\alpha^2\beta^2+2\beta^4}{4096\pi ^4x_{f_1}}\\&+\frac{\alpha^4-4\alpha^2\beta^2+2\beta^4}{2048\pi ^4}\eta_{-2/3}l_p+\frac{\alpha^4-4\alpha^2\beta^2+2\beta^4}{1024\pi ^4}\eta_{-2/3}l_p\ln x_{f_1}~.
\end{split}
\end{equation}
\end{widetext}

 \subsection{Energy output as a function of time for $\omega_q=-1/3$}
\noindent We shall now calculate energy radiation rate with time for $\omega_{q}= -1/3 $ case. For $\omega_q=-1/3$ the evaporation equation reads
\begin{equation}\label{4.26}
\frac{1-\eta_{-1/3}}{2}\frac{dr_H}{dt}=-4\pi \sigma r_H^2T_H^4~.
\end{equation}
Following the procedure in the earlier subsection, we can recast the above equation as
\begin{equation}\label{4.27}
\begin{split}
(1-\eta_{-1/3})\left(x^2+\bar{C}_1x+\bar{C}_0+\frac{\bar{C}_{-1}}{x}+\frac{\bar{C}_{-2}}{x^2}\right)\frac{dx}{dt}=-\frac{1}{t_{Ch.}}
\end{split}
\end{equation}
where the constant parameters
\begin{equation}\label{4.28}
\begin{split}
\bar{C}_1&=\frac{\alpha}{2\pi}~,\\
\bar{C}_0&=\frac{3\alpha^2-2\beta^2}{32\pi^2}~,\\\bar{C}_{-1}&=\frac{\alpha (\alpha^2-2\beta^2)}{128\pi^3}~,\\
\bar{C}_{-2}&=\frac{\alpha^4-4\alpha^2\beta^2+2\beta^4}{4096\pi^4}~.
\end{split}
\end{equation}
After integrating eq.(\ref{4.27}), we can recast it as
\begin{widetext}
\begin{equation}\label{4.29}
\begin{split}
&(1-\eta_{-1/3})\left(\frac{x^3}{3}+\frac{\bar{C}_1x^2}{2}+\bar{C}_0x+\bar{C}_{-1}\ln x-\frac{\bar{C}_{-2}}{x}\right)+A_1=-\frac{t}{t_{\text{ch}}}~.
\end{split}
\end{equation}
Again setting $t=0$, we can obtain the form of $A_1$
\begin{equation}\label{4.30}
\begin{split}
A_1&=-(1-\eta_{-1/3})\Bigr(\frac{x_i^3}{3}+\frac{\bar{C}_1x_i^2}{2}+\bar{C}_0x_i+\bar{C}_{-1}\ln x_i-\frac{\bar{C}_{-2}}{x_i}\Bigr)~.
\end{split}
\end{equation}
\end{widetext}
Following the perturbative approach used in the previous subsection, we can obtain the zeroth order as well as the $\eta$ order solution to be 
\begin{align}
x_0&=\left(x_i^3-\frac{3t}{t_{\text{ch}}}\right)^{\frac{1}{3}}~,\label{4.31}\\
x_\eta&=-\frac{\eta_{-1/3}t}{x_0^2t_{\text{ch}}}~.\label{4.32}
\end{align}
\begin{figure}[ht!]
\begin{center}
\includegraphics[scale=0.28]{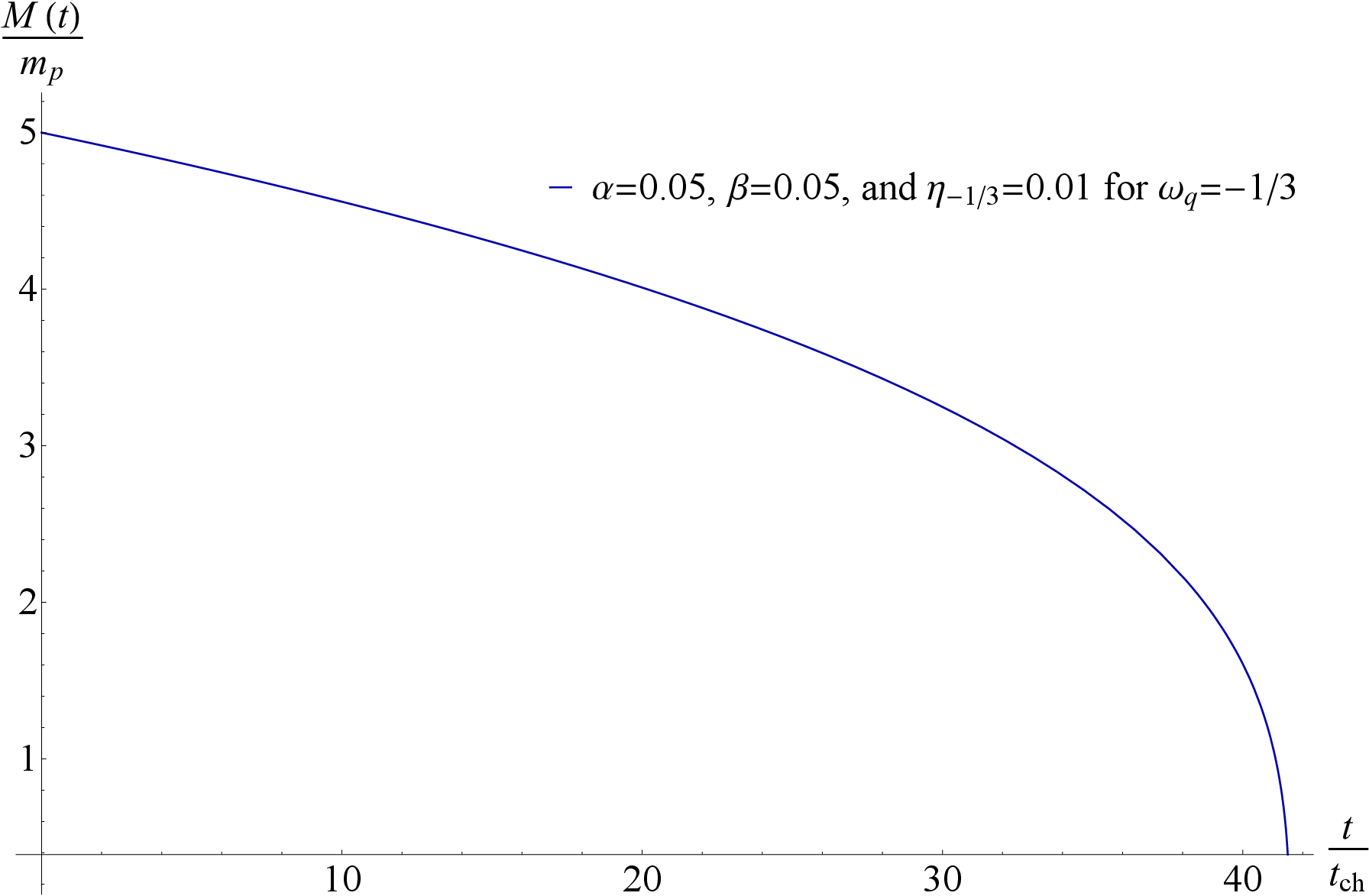}
\caption{$x(t)=\frac{M(t)}{m_p}$ is plotted against the dimensionless time $\frac{t}{t_{\text{ch}}}$ for $\alpha=\beta=0.05$ and $\eta_{-1/3}=0.01$\label{Fig12}.}
\end{center}
\end{figure}
It is easy to check that the zeroth order solution is the same for the $\omega_q=-2/3$ and the $\omega_q=-1/3$ case. Again substituting first $x=x_0+x_\eta+x_1$ and then $x=x_0+x_\eta+x_1+x_2$, we can obtain perturbatively the first and second order solution to be 
\begin{widetext}
\begin{align}
x_1&=-\frac{\alpha}{4\pi x_0^2}(x_0^2-x_i^2)+\frac{\alpha \eta_{-1/3}tx_i^2}{2\pi t_{\text{ch}}x_0^6}~,\label{4.33}\\
x_2&=\frac{\alpha^2}{16\pi^2 x_0^5}(x_0^4-x_i^4)-\frac{3\alpha_2-2\beta^2}{32\pi^2x_0^2}(x_0-x_i)-\frac{\eta_{-1/3}t}{t_{\text{ch}}}\frac{(x_0^4(\alpha^2-2\beta^2)-2x_ix_0^3(3\alpha^2-2\beta^2+10x_i^4\alpha^2))}{32\pi^2x_0^8}~.\label{4.34}
\end{align}
\end{widetext}
With the solution of $x$ in hand, we shall now plot $x(t)$ against $\frac{t}{t_{\text{ch}}}$ in Fig.(\ref{Fig12}).
As has been observed in the earlier case, we observe similar behaviour of the mass evaporation of the black hole with time in the $\omega_q=-1$ case. We observe that the mass starts to decay faster with increasing time.

\noindent Now, for $\omega_q=-1/3$, at $t=t_\mathcal{E}$, $x_f=\frac{1-\eta_{-1/3}}{8\pi}(2\beta-\alpha)$ and as a result one can obtain the evaporation time by integrating eq.(\ref{4.27}) from $x_i$ to $x_f$ as
\begin{widetext}
\begin{equation}\label{4.35}
\begin{split}
&\frac{t_{\mathcal{E}}}{t_{\text{ch}}}=(1-\eta_{-1/3})\Bigr(\frac{x_i^3-x_f^3}{3}+\frac{\bar{C}_1(x_i^2-x_f^2)}{2}+\bar{C}_0(x_i-x_f)+\bar{C}_{-1}\ln \left(\frac{x_i}{x_f}\right)-\bar{C}_{-2}\left(\frac{1}{x_i}-\frac{1}{x_f}\right)\Bigr)
\end{split}
\end{equation}
from which one can write down the expanded form of the evaporation time up to $\mathcal{O}(\eta_{-1/3}\alpha^4,\eta_{-1/3}\alpha^2\beta^2,\eta_{-1/3}\beta^4)$ as
\begin{equation}\label{4.36}
\begin{split}
\frac{t_{\mathcal{E}}}{t_{\text{Ch}}}&\simeq(1-\eta_{-1/3})\Bigr(\frac{x_i^3}{3}+\frac{\bar{C}_1x_i^2}{2}+\bar{C}_0x_i+\bar{C}_{-1}\ln x_i-\frac{C_{-2}}{x_i}\Bigr)-\frac{1-4\eta_{-1/3}}{3}x_{f_1}^3-\frac{(1-3\eta_{-1/3})\bar{C}_1}{2}x_{f_1}^2\\&-(1-2\eta_{-1/3})\bar{C}_0x_{f_1}+\eta_{-1/3}\bar{C}_{-1}-(1-\eta_{-1/3})\bar{C}_{-1}\ln x_{f_1}+\frac{\bar{C}_{-2}}{x_{f_1}}
\end{split}
\end{equation}
\end{widetext}
where $x_{f_1}=\frac{2\beta-\alpha}{8\pi}$ gives the final value of $x$ when the evaporation process stops in the absence of any quintessential energy matter surrounding the black hole.
 \subsection{Energy output as a function of time for $\omega_q=-1$}
\noindent We shall now write down the evaporation equation corresponding to the $\omega_q=-1$ as
\begin{equation}\label{4.37}
\begin{split}
\frac{1-3\eta_{-1}r_H^2}{2}\frac{dr_H}{dt}=-4\pi\sigma r_H^2T_H^4~.
\end{split}
\end{equation}
Again by using $r_H=2l_p x$, we can write down the first-order non-linear differential equation in $x$ as
\begin{widetext}
\begin{equation}\label{4.38}
\begin{split}
&\Bigr(x^2+\tilde{C}_4x^4+\tilde{C}_3x^3+\tilde{C}_2x^2+\tilde{C}_1x  i+\tilde{C}_0+\frac{\tilde{C}_{-1}}{x}+\frac{\tilde{C}_{-2}}{x^2}\Bigr)\frac{dx}{dt}=-\frac{1}{t_{\text{ch}}}~,
\end{split}
\end{equation}\end{widetext}
where the constants are given as
\begin{equation}\label{4.39}
\begin{split}
\tilde{C}_4&=-12\eta_{-1}m_p^2~,\\
\tilde{C}_3&=-\frac{6\eta_{-1}m_p^2\alpha}{\pi}~,\\
\tilde{C}_2&=-\frac{3\eta_{-1}m_p^2(3\alpha^2-2\beta^2)}{8\pi^2}~,\\
\tilde{C}_1&=\frac{\alpha}{2\pi}-\frac{3\eta_{-1}m_p^2\alpha (\alpha^2-2\beta^2)}{32\pi^3}~,\\
\tilde{C}_0&=\frac{3\alpha^2-2\beta^2}{32\pi^2}-\frac{3\eta_{-1}m_p^2(\alpha^4-4\alpha^2\beta^2+2\beta^4)}{1024\pi^4}~,\\
\tilde{C}_{-1}&=\frac{\alpha(\alpha^2-2\beta^2)}{128\pi^3}~,\\
\tilde{C}_{-2}&=\frac{\alpha^4-4\alpha^2\beta^2+2\beta^4}{4096\pi^4}~.
\end{split}
\end{equation}

\begin{figure}[ht!]
\begin{center}
\includegraphics[scale=0.28]{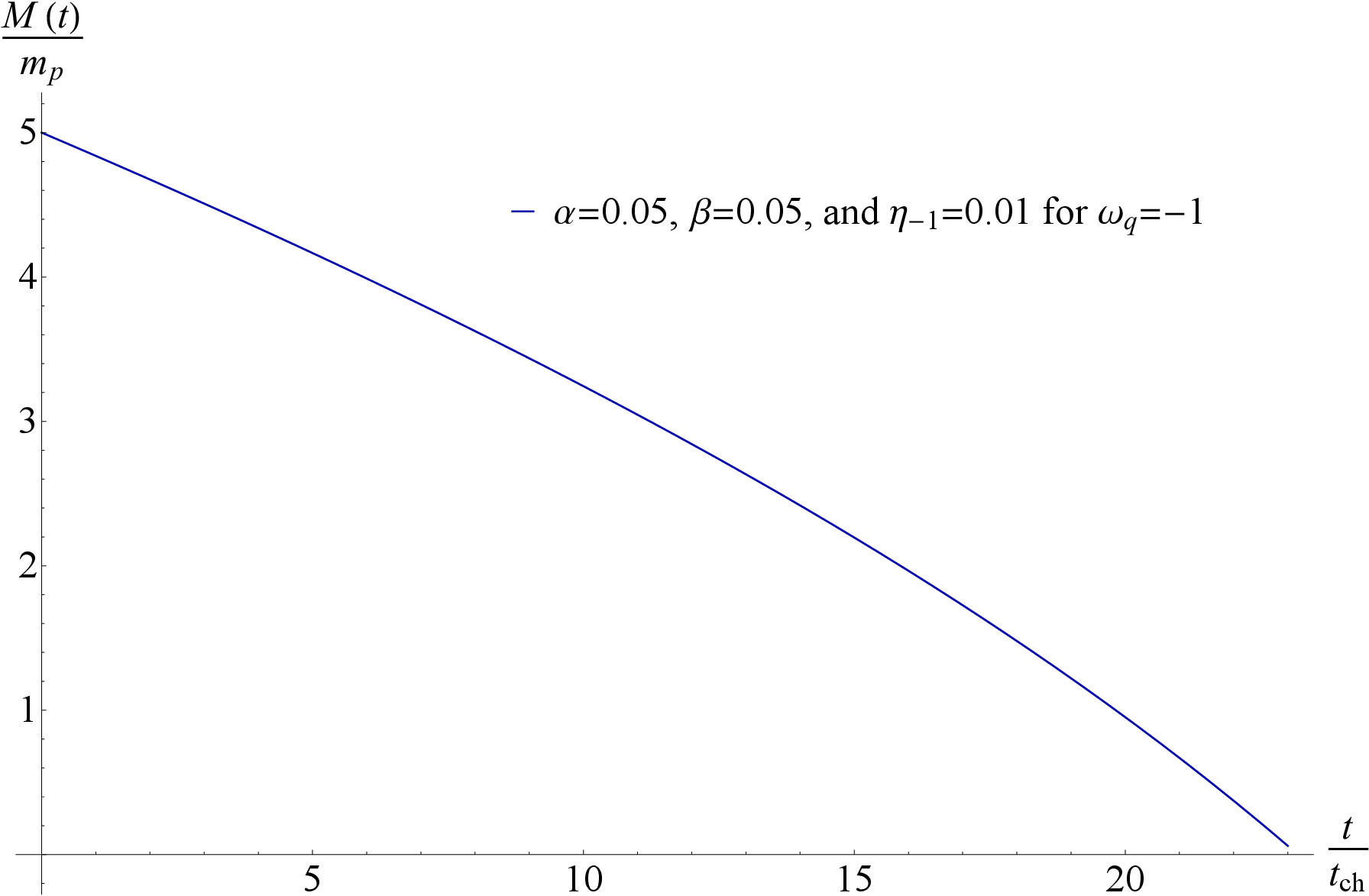}
\caption{$x(t)=\frac{M(t)}{m_p}$ is plotted against the dimensionless time $\frac{t}{t_{\text{ch}}}$ for $\alpha=\beta=0.05$ and $\eta_{-1}=0.01$\label{Fig13}.}
\end{center}
\end{figure}

\noindent Integrating eq.(\ref{4.38}) one can obtain the evaporation equation of the black hole
\begin{equation}\label{4.40}
\begin{split}
&\frac{x^3}{3}+\frac{\tilde{C}_4x^5}{5}+\frac{\tilde{C}_3x^4}{4}+\frac{\tilde{C}_2x^3}{3}+\frac{\tilde{C}_1x^2}{2}+\tilde{C}_0x+\tilde{C}_{-1}\ln x\\&-\frac{\tilde{C}_{-2}}{x}+A_2=-\frac{t}{t_{\text{ch}}}~,
\end{split}
\end{equation}
where the constant $A_2$ can be determined by using the initial time condition as
\begin{equation}\label{4.41}
\begin{split}
A_2&=-\frac{x_i^3}{3}-\frac{\tilde{C}_4x_i^5}{5}-\frac{\tilde{C}_3x_i^4}{4}-\frac{\tilde{C}_2x_i^3}{3}-\frac{\tilde{C}_1x_i^2}{2}-\tilde{C}_0x_i\\&-\tilde{C}_{-1}\ln x_i+\frac{\tilde{C}_{-2}}{x_i}~.
\end{split}
\end{equation}
We shall again follow a perturbative approach to obtain the correct mass time relation.
The zeroth order solution comes out to be the same as the earlier two cases. Following the perturbative approach the $\eta$ order solution takes the form
\begin{equation}\label{4.42}
x_\eta=\frac{12 l_p^2\eta_{-1}}{5x_0^2}(x_0^5-xi^5)~.
\end{equation}
Next, as before, one can obtain the first order contribution in the GUP parameter to $x$ as
\begin{equation}
\begin{split}
x_1&=-\frac{\alpha}{4\pi x_0^2}(x_0^2-x_i^2)\\&-\frac{3\alpha \eta_{-1}l_p^2}{10\pi x_0^5}(5x_0^7-6x_i^2x_0^5+5x_i^4x_0^3-4x_i^7)\label{4.43}~.
\end{split}
\end{equation}
One can again use $x=x_0+x_\eta+x_1+x_2$ and obtain the form of the second order contribution in the GUP parameter to $x$ to be
\begin{widetext}
\begin{equation}\label{4.44}
\begin{split}
x_2=&\frac{\alpha^2}{16\pi^2 x_0^5}(x_0^4-x_i^4)-\frac{3\alpha_2-2\beta^2}{32\pi^2x_0^2}(x_0-x_i)+\frac{\eta_{-1}l_p^2}{40\pi^2x_0^8}\bigr[x_0^9(3\alpha^2+14\beta^2)+9x_0^8x_i
(3\alpha^2-2\beta^2)-30x_i^2x_0^7\alpha^2\\-&5x_i^3x_0^6(3\alpha^2-2\beta^2)-3x_i^5x_0^4(\alpha^2-
2\beta^2)+12x_i^6x_0^3(4\alpha^2-\beta^2)-30x_i^9\alpha^2\bigr]~.
\end{split}
\end{equation}
\end{widetext}
We shall now again plot $x$ versus $\frac{t}{t_{\text{ch}}}$ in Fig.(\ref{Fig13}) for $\alpha=\beta=0.05$ and $\eta_{-1}=0.01$. 
We observe an almost similar behaviour corresponding to the earlier two cases. However, it is important to observe that the increase in the decay rate is no longer very apt from the Figure itself compared to the earlier cases.

\noindent Now the initial mass of the black hole is $2m_p x_i$ whereas the final mass at which the evaporation process stops is given by its remnant mass $M_f=\frac{m_p}{8 \pi }(2\beta - \alpha)-\frac{\eta_{-1}m_p^3}{128\pi^3}(2\beta-\alpha)^3$. The final integration limit reads
\begin{equation}\label{4.45}
x_f=\frac{2\beta - \alpha}{8 \pi }-\frac{\eta_{-1}m_p^2}{128\pi^3}(2\beta - \alpha)^3~.
\end{equation}
Following earlier analysis one can obtain the evaporation time upto $\mathcal{O}(\eta_{-1}\alpha^4,\eta_{-1}\alpha^2\beta^2,\eta_{-1}\beta^4)$ to be
\begin{widetext}
\begin{equation}\label{4.46}
\begin{split}
\frac{t_{\mathcal{E}}}{t_{\text{ch}}}\simeq \frac{x_i^3}{3}-\frac{x_{f_1}^3}{3}+\frac{\tilde{C}_4x_i^5}{5}+\frac{\tilde{C}_3x_i^4}{4}+\frac{\tilde{C}_2x_i^3}{3}+\frac{\tilde{C}_1x_i^2}{2}-\frac{\alpha}{4\pi}x_{f_1}^2+\tilde{C}_0x_i-\frac{3\alpha^2-2\beta^2}{32\pi^2}x_{f_1}+\tilde{C}_{-1}\ln \left[\frac{x_i}{x_{f_1}}\right]-\tilde{C}_{-2}\left[\frac{1}{x_i}-\frac{1}{x_{f_1}}\right]~,
\end{split}
\end{equation}
\end{widetext}
where $x_{f_1}=\frac{2\beta-\alpha}{8\pi}$.
\section{A comparative study with other woks in the literature}\label{S4B}
\noindent Here, we discuss the key differences of our analysis with some of the existing works on black hole thermodynamics in the generalized uncertainty principle framework. As we have already highlighted throughout the manuscript, the key differences between \cite{QuintSchwarzQGUP} and our current analysis, lies in the fact that the form of the generalized uncertainty principle is more general than the one investigated in \cite{QuintSchwarzQGUP}. One important aspect of using a GUP framework is that the black hole does not evaporante completely. We shall now look at few works on black hole thermodynamics \cite{DuttaPadhyay,BHThermo8,BHThermo10,BHThermo11,
BHThermo16} and do a side by side comparison. In \cite{DuttaPadhyay}, the authors of investigated black hole thermodynamics in the simple quadratic generalized uncertainty principle framework for a Schwarzschild and Reissner-Nordstr\"{o}m black hole and obtained the mass-temperature relation, specific heat and enetropy. In our analysis, we have considered the LQGUP framework and the Schwarzschild black hole is surrounded by quintessence matter. Hence, the lapse function is different than the case considered in \cite{DuttaPadhyay}. Similarly in \cite{BHThermo11,BHThermo16}, not only the GUP structure is different from the one used in this current analysis, but also the black hole metric does not have any contribution from quintessence matter. Finally, in \cite{BHThermo8,BHThermo10}, the authors have made use of the LQGUP framework and obtained the entropy, evaporation time, and modified dispersion relation for different black hole spacetime. The primary difference in our analysis lies in the fact that the black hole spacetime is surrounded by quintessence matter. Another important difference of this approach with most of the previous works is that the small parameters in this work have been treated perturbatively and the orders in the GUP parameters have been taken consistently in obtaining the results\footnote{This section has been added to make a comparative study with our work with the previous works in the literature.}.
\section{Conclusion}\label{S5}
\noindent In this analysis we have considered a Schwarzschild black hole surrounded by quintessence matter in the linear quadratic generalized uncertainty principle framework. In this work, we have extended the analysis presented in \cite{QuintSchwarzQGUP} by considering linear order correction in the momentum uncertainty in the well-known form of the generalized uncertainty principle. We start by calculating the critical mass and the specific heat of the black hole. We then plot the specific heat against the mass and then against the event horizon radius of the black hole. From Fig.(\ref{Fig1}), we observe that the approximate solutions are quite appropriate for black holes with smaller masses. Then in Fig.(s)(\ref{Fig2},\ref{Fig3},\ref{Fig5}), we have compared the specific heat for the quadratic generalized uncertainty principle case against the linear and quadratic generalized uncertainty principle case and observed that the specific heat has a higher rate of descent and ascent for the linear and quadratic generalized uncertainty principle case than the quadratic generalized uncertainty principle case. Next, in Fig.(\ref{Fig6}), we have plotted the specific heat of the black hole against the event horizon radius for different values of the quintessence parameter. We observe that for  higher values of the quintessence parameter, the specific heat has a steeper decent than the other two cases with $\omega_q=-\frac{2}{3}$ and $\omega_q=-1$ when the event horizon radius of the black hole keeps on increasing. We also observe that for the lowest value of the quintessence parameter ($\omega_1=-1$), the specific heat attains a minima for a certain value of the event horizon radius and then it keeps on increasing with increasing values of $r_H$. We have then computed the remnant mass of the black hole and have found it to be the same as the critical mass of the black hole. Then using the specific heat value of the black hole corresponding to three fixed values of the quintessential state parameter, we investigate the form of the Bekenstein-Hawking black hole entropy of the black hole. We observe sub-leading logarithmic corrections after the leading order ``\textit{area by four}" term in the entropy. It is important to observe that the logarithmic contribution is coming solely due to the quadratic GUP parameter ($\beta$) and has no connection with the linear parameter whereas the linear parameter contributes mainly in very small fractional area contributions to the entropy. Next, we calculated the energy density using the entropy results for the three cases and plotted the $\omega_q=-2/3$ and $\omega_q=-1/3$ cases against the event horizon radius of the black hole. We specifically observe that the energy density is a constant, independent of the horizon radius, for the $\omega_q=-1$ case. Finally, in section (\ref{S4}), we have expressed the energy output of the black hole as a function of time for the three distinct cases with different fixed constant values of $\omega_q$. We use the Stefan-Boltzmann law to obtain the evaporation equation of the black hole and from there, we have obtained the mass-time relation for the black hole. Finally, from the differential equation of the mass of the black hole, we have obtained the evaporation time of the black hole by integrating from the initial mass of the black hole to its remnant mass. Then we plotted $x(t)=M(t)/m_p$ against the dimensionless time $t/t_{\text{ch}}$ to investigate the time dependence of the mass of the black hole for the three cases. 
\section*{Acknowledgement}
\noindent We thank the anonymous referee for constructive criticism of our paper.

\end{document}